\documentclass[final,3p,times]{elsarticle}
\usepackage{graphicx}% Include figure files
\usepackage{dcolumn}% Align table columns on decimal point
\usepackage{bm}% bold math
\usepackage{epsfig}
\usepackage{subfigure}
\usepackage{graphics}
\usepackage{amssymb}
\usepackage{amsthm}
\usepackage{amsmath}
\usepackage{array}
\usepackage{tabularx}
\usepackage{multirow}
\usepackage{amsmath}
\usepackage{xcolor}
\usepackage{ulem}
\usepackage{extarrows}

\usepackage{caption}
\usepackage{algorithm}
\usepackage{algpseudocode}
\usepackage{amsthm} % Remark
\newtheorem*{remark}{Remark}

\graphicspath{{./pic/}}

\begin{document}
%\pagecolor{ultramarine!70} %page color

\begin{frontmatter}
\title{A third-order discrete unified gas kinetic scheme for continuum and rarefied flows: low-speed isothermal case}

\author[add1]{Chen Wu}
%\affiliation{State Key Laboratory of Coal Combustion, Huazhong
%University of Science and Technology, Wuhan 430074, China}

\author[add2]{Chang Shu}

\author[add1,add3,add4]{Baochang Shi \corref{mycorrespondingauthor}}\cortext[mycorrespondingauthor]{Corresponding author}
\textbf{\ead{shibc@hust.edu.cn}}

\author[add2]{Zhen Chen}

%\email[Corresponding author]{(shibc@hust.edu.cn)}
%\affiliation{School of Mathematics and Statistics, Huazhong
%University of Science and Technology, Wuhan 430074, China}%

%\author[mysecondaryaddress]{Zhenhua Chai}\cortext[mycorrespondingauthor]{Corresponding author}
%\ead{hustczh@hust.edu.cn}
%\affiliation{School of Mathematics and Statistics, Huazhong
%University of Science and Technology, Wuhan 430074, China}%

\address[add1]{State Key Laboratory of Coal Combustion, Huazhong
University of Science and Technology
Wuhan, 430074, China}
\address[add2]{Department of Mechanical Engineering, National University of Singapore, 10 Kent Ridge Crescent, Singapore 119260, Singapore}
\address[add3]{School of Mathematics and Statistics, Huazhong
	University of Science and Technology, Wuhan 430074, China}
\address[add4]{Hubei Key Laboratory of Engineering Modeling and Scientific Computing, Huazhong University of Science and Technology, Wuhan
430074, China}

\date{\today}% It is always \today, today,

%%%%% Begin Abstract %%%%%%%%%%%
\begin{abstract}
An efficient third-order discrete unified gas kinetic scheme (DUGKS) with efficiency is presented in this work for simulating continuum and rarefied flows. By employing two-stage time-stepping scheme and the high-order DUGKS flux reconstruction strategy, third-order of accuracy in both time and space can be achieved in the present method. It is also analytically proven that the second-order DUGKS is a special case of the present method. Compared with the high-order lattice Boltzmann equation {LBE} based methods, the present method is capable to deal with the rarefied flows by adopting the Newton-Cotes quadrature to approximate the integrals of moments. Instead of being constrained by the second-order (or lower-order) of accuracy in time splitting scheme as in the conventional high-order Runge-Kutta (RK) based kinetic methods, the present method solves the original BE, which overcomes the limitation in time accuracy. Typical benchmark tests are carried out for comprehensive evaluation of the present method. It is observed in the tests that the present method is advantageous over the original DUGKS in accuracy and capturing delicate flow structures. Moreover, the efficiency of the present third-order method is also shown in simulating rarefied flows.

\end{abstract}
%%%%% end %%%%%%%%%%%

%\keywords{DUGKS, incompressible, finite-volume}


\end{frontmatter}

%\maketitle

\section{\label{sec:level1}Introduction}
\label{sec:introduction}
In recent years, the kinetic theory based numerical methods for simulating multiscale fluid flows have attracted much attention \cite{succi2001lattice, d2002multiple, fan2001statistical, xu2001gas, xu2010unified, guo2013discrete, shu2014development, yan2013successive, dimarco2011exponential, yang2016numerical}. Because of their wide and valuable applications in many frontier scientific fields such as spacecraft \cite{li2009gas, chen2012unified} and microelectromechanical systems \cite{fan2001statistical}, higher demands in accuracy and efficiency is raised, due to which the development of high-order kinetic scheme is necessary.

Various second-order schemes have been developed within the kinetic framework, including the lattice Boltzmann equation (LBE) based method \cite{succi2001lattice, d2002multiple, du2006multi, guo2008lattice}, the gas kinetic scheme (GKS) \cite{xu2001gas, xu2010unified, guo2013discrete, chen2012unified, wu2016discrete}, and the Boltzmann equation (BE) based discrete velocity method (DVM) \cite{yan2013successive, dimarco2011exponential, yang2016numerical}. However, efficient higher-order kinetic methods are sparse. For BE based methods, improving the spatial accuracy is relatively easier, which is usually realized by  implementing high-order discretization scheme in space. The major challenge in developing higher-order method is the stiffness of the collision operator in BE \cite{dimarco2011exponential} during the time evolution. To overcome this problem, most high-order numerical methods for BE are based on a time splitting scheme \cite{bobylev2001error, ohwada1998higher, dellar2013interpretation}, which split BE into two separate equations: the transport equation and collision equation. Then, a variety of high-order numerical schemes can be used to solve the stiff collision equation, such as implicit-explicit Runge-Kutta (IMEX) scheme \cite{hejranfar2017high}, exponential Runge-Kutta scheme \cite{dimarco2011exponential}, penalty-based implicit scheme \cite{yan2013successive} and the variation of time relaxed (TR) scheme \cite{filbet2003high} et al. However, most time splitting schemes only have the second or even lower order of accuracy in its splitting process \cite{issa1986solution, sportisse2000analysis}, which encumbers the improvement of the overall order of accuracy. For example, the two high-order methods proposed by Hejranfar et al. \cite{hejranfar2017high} and Su et al. \cite{su2015parallel} use the high-order Runger-Kutta scheme \cite{hejranfar2017high, shu1988efficient} and high-order weighted essentially non-oscillatory finite-difference (WENO) \cite{jiang1996efficient} or Galerkin scheme \cite{cockburn1989tvb} to treat the temporal and spatial terms in split BE respectively which provide more accurate results than the second order methods. But the third or higher order of accuracy could not be validated evidently. This is induced by the incompatible orders of accuracy between the splitting scheme and discretization scheme. To resolve this issue, higher-order time-splitting schemes have been reported \cite{geiser2008iterative, cox2002exponential, kassam2005fourth}, such as the exponential time differencing (ETD) Runge-Kutta method. But it is too complex to use in practice, especially in simulating multidimensional fluid flows. On the other hand, for LBE based numerical methods, the order of accuracy is also limited no more than the second order. The reason is that the LBE is obtained by integrating BE along the characteristic line with the second order approximation \cite{succi2001lattice, du2006multi}. Meanwhile, the LBE methods can hardly adopt the Newton-Cotes quadrature, which implies their difficulty in simulating highly non-equilibrium flows is also difficult. Thus, other ways should be explored to improve the accuracy order of the numerical methods for BE.

A novel two-stage fourth order time-accurate discretization (TFTD) scheme is proposed by Li et al. \cite{li2016two}, which has already been applied to generalized Riemann problem solver (TFTD-GPR) and GKS solver (TFTD-GKS) \cite{pan2016efficient} for Navier-Stokes (NS) equations successfully. Different from the Runge-Kutta methods, the key point of TFTD is that both of the flux function and its time derivative are used in this method. However, the TFTD scheme is developed for the hyperbolic conservation laws \cite{li2016two}, which cannot be used to BE directly. The first reason is that the existence of collision term in BE prevents the implementation of Lax-Wendroff type time stepping scheme, and it also brings the difficulties in solving the flux function and its time derivative. The second reason is that, unlike TFTD-GKS in which the time accurate flux function for the conservative flow variables of NS equations can be obtained by kinetic theory, it is hard to adopt such strategy to construct a flux function for the particle distribution functions (PDFs) of BE. It is noteworthy that, in the proving process of TFTD, if we only consider the third-order of accuracy rather than the forth-order, the system of equations is no longer unique, which endows us the flexibility of constructing a series of third-order schemes by adjusting the free parameters. This feature provides possibilities to develop a third-order time evolution scheme for BE, which only needs to solve the flux at intermediate time steps. At the same time, the discrete unified gas kinetic scheme (DUGKS) can provide high-order numerical solutions of the intermediate flux function. 

DUGKS is a second-order finite volume method proposed by Guo et. al. \cite{guo2013discrete, guo2015discrete} which is efficient for all Knudsen number flows and combines the advantages of LBE and unified GKS (UGKS). The reason for the success of DUGKS is employing a LBE based transformation of distribution function with collision effect to simplify the evaluation of the flux at a cell interface. Besides, the application of midpoint and trapezoidal rule for time discretization also ensures the second-order accuracy. However, due to the coupling between the evolution time step and the PDF streaming distance in the second-order flux reconstruction process, it is difficult to improve the time accuracy of DUGKS by the high-order Runge-Kutta scheme. 

In current work, a two-stage third-order time-accurate DUGKS for BE is developed with efficiency for simulating continuum and rarefied flows. With the requirement of the third-order time-accuracy, the analysis of BE using TFTD provides a series of time discretization schemes depending on free parameters. Considering the requirement in accuracy and efficiency, we can specify a set of the parameters and determine the proper two-stage third-order time discretization schemes for BE. Finally, the present method can be constructed by discretizing BE in space by employing the finite volume method (FVM) and a higher-order DUGKS reconstruction method. In the derivation process, the original second-order DUGKS can be proven to be a special case of the present method during the analyzing. Different from many other higher-order methods for BE \cite{ohwada1998higher,  hejranfar2017high, su2015parallel}, the third-order accuracy of the present method can be validated by numerical test case exactly. And compared with the high-order LBE-based methods \cite{hejranfar2017high, meng2011accuracy, kim2008accuracy}, the present method is capable of dealing with the rarefied flows by adopting the Newton-Cotes quadrature to approximate the integrals of moments. Besides, compared with the original second-order DUGKS \cite{guo2013discrete}, the present method performs better in predicting delicate flow structures and providing accurate time-depending solutions. It is also revealed in the comparison of computational efforts that the present third-order method consumes almost the same computational time and virtual memories to get comparable results. Compared with other high-order methods for BE in continuum regime \cite{hejranfar2017high, su2015parallel}, the present method also shows its advantage in numerical stability, which is reflected as the larger stability boundary in time step.

The paper is organized as follows. Section \ref{sec:numerical_methods} presents the derivation of the proposed method as well as its boundary treatment. In Section \ref{sec:numerical_cases}, four representative numerical cases for continuum and rarefied flows are tested to validate the present method. Section \ref{sec:conclusions} draws the conclusion.

\section{Numerical methods} 
\label{sec:numerical_methods}

\subsection{Third-order DUGKS}
In this section, the third-order DUGKS will be discussed from two aspects separately: time discretization and space discretization. The TFTD method are used to construct time marching scheme of BE which fulfills the requirements in accuracy and efficiency. Then the FVM is employed to discretize the BE in space, in which a higher-order DUGKS reconstruction scheme is proposed to calculate the micro-flux.

\subsubsection{Time discretization}
\label{sec:time_accuracy}
%TFTD \cite{li2016two} is a newly proposed high-order time-accurate approach. However, it is hard to solve BE using a Lax-Wendroff type solver without assumptions or simplifications.

Consider the time-dependent Boltzmann equation,
\begin{equation}
\label{BE}
	\frac{\partial f}{\partial t} = L(f) + \Omega(f) := M(f),
\end{equation}
where the transport term $L$ and the collision term $\Omega$ can be expressed as
\begin{equation}
L(f) = -\bm{\xi}  \cdot \nabla f,
\end{equation} 
\begin{equation}
\Omega (f) =  - \frac{1}{\tau }\left[ f  - f^{eq} \right],
\end{equation}
where $f = f({\bm{x}},\bm{\xi} ,t)$ is the particle distribution function associated with the particle velocity $\bm{\xi}$, location $\bm{x}$ and time $t$; $\tau$ is the relaxation time; and $f^{eq}$ is the Maxwellian equilibrium distribution function in the form of
\begin{equation}
\label{maxwellian_expansion}
{f^{eq}} = \frac{\rho }{{{{(2\pi RT)}^{D/2}}}}\exp \left(  - \frac{{{\left| {\bm{\xi}  - \bm{u}} \right|}^2}}{{2RT}} \right),
\end{equation}
where $\rho$ is the density, $R$ is the gas constant, $T$ is the temperature, $D$ is the spatial dimension, $\bm{u}$ is the velocity of fluid. 
In low Mach number cases $(
{Ma}\mathrm{\approx}\frac{\left|{\bm{u}}\right|}{\sqrt{RT}}\mathrm{\ll}{0}{\mathrm{.}}{1}
 )$, the second-order approximation of Maxwellian equilibrium distribution function can be derived from its Hermite or Taylor expansions \cite{guo2013discrete} as 
\begin{equation}
{f}^{eq}{{=}}{\frac{\mathrm{\rho}}{{(}{2}{\pi}{RT}{)}^{D/2}}}\exp \left ({\mathrm{{-}}{\frac{{\left|{\bm{\xi}}\right|}^{2}}{2RT}}}\right) \left[{{1}\mathrm{{+}}{\frac{\bm{\xi}\cdot{\mathbf{u}}}{RT}}\mathrm{{+}}{\frac{{(}\bm{\xi}\cdot{\mathbf{u}}{)}^{2}}{2(RT{)}^{2}}}\mathrm{{-}}{\frac{{\left|{\mathbf{u}}\right|}^{2}}{2RT}}}\right].
\end{equation}
The conservative flow variables can be obtained from the moments of the distribution function,
\begin{equation}
{W}\mathrm{{=}}\left({\begin{array}{c}{\mathit{\rho}}\\{\mathit{\rho}{\bm{u}}}\end{array}}\right)\mathrm{{=}}\int{\mathit{\bm{\psi}}{\mathrm{(}}\mathit{\bm{\xi}}{\mathrm{)}}{fd}\mathit{\bm{\xi}}{\mathrm{,}}}
\end{equation}
where $\bm{\psi} = (1, \bm{\xi})^T$ is the collision invariant.

Integrating Eq. \eqref{BE} over the time interval $[t_n,t_n+\Delta t]$ gives
\begin{equation}
		f^{n+1} = f^n + \int_{t_n}^{t_n+\Delta t}M[f(t)]dt, \\
	\label{feqIntC}
\end{equation}
where $f^n = f(\bm{x}, \bm{\xi}, t_n)$.

According to Li's work \cite{li2016two}, performing the chain rule on $\frac{\partial }{{\partial t}}M [f(t)]$ yields
\begin{equation}
\frac{\partial }{{\partial t}}M [f(t)] = \frac{\partial }{{\partial f}}M [f(t)] M [f(t)].
\end{equation}
And the Taylor series expansion of the time integration term in Eq. \eqref{feqIntC} at $f_n$ gives obtained as
\begin{equation}
	\begin{split}
		\int_{t_n}^{t_n+\Delta t}M(f)dt 
		= &\Delta tM({f^n}) + \frac{{\Delta {t^2}}}{2}{M_f}({f^n})M({f^n}) + \frac{{\Delta {t^3}}}{6}\left[ {M_{f}^2({f^n})M({f^n}) + {M_{ff}}({f^n}){M^2}({f^n})} \right] + O({\Delta {t^4}}), \\
	\end{split}
	\label{ExpansionOri}
\end{equation}
where ${M_f} = \frac{\partial }{{\partial f}}M$.

Then, we want to construct a solution for Eq. \eqref{feqIntC} using the intermediate values $f^* = f(\bm{x}, \bm{\xi}, t_*)$ at time $t_* = t_n+A\Delta t$ and $f^{**} = f(\bm{x}, \bm{\xi}, t_{**})$  at time $t_{**} = t_n+K\Delta t$, where $0 \le A < K < 1$, to achieve the third order time accuracy. Performing Taylor series analysis of the intermediate values leads to
\begin{subequations}
	\begin{equation}
	f^*=f^n + A \Delta t M(f^n) + \frac{1}{2} A^2 \Delta t^2 \frac{\partial}{\partial t} M(f^n).
	\end{equation}
	\begin{equation}
	{f^{**}} = {f^n} + K\Delta tM({f^n}) + \frac{1}{2}{K^2}\Delta {t^2}\frac{\partial }{{\partial t}}M({f^n}).
	\end{equation}
\end{subequations}
And the solution can be set as
\begin{equation}
\begin{split}
{f^{n + 1}}{\rm{ }} = {f^n} &+ {\rm{ }}\Delta t\left( {{B_0}M({f^n}) + {B_1}M({f^*}) + {B_2}M({f^{**}}) + {B_3}M({f^{n + 1}})} \right) \\ &+ \frac{1}{2}\Delta {t^2}\left( {{C_0}\frac{\partial }{{\partial t}}M({f^n}) + {C_1}\frac{\partial }{{\partial t}}M({f^*}) + {C_2}\frac{\partial }{{\partial t}}M({f^{**}}) + {C_3}\frac{\partial }{{\partial t}}M({f^{n + 1}})} \right),
\end{split}
\label{EvolutionEquationFirst_M}
\end{equation}
or
\begin{equation}
\footnotesize
\begin{split}
{f^{n + 1}}{\rm{ }} = {f^n} &+ {\rm{ }}\Delta t\left( {{B_0}L({f^n}) + {B_1}L({f^*}) + {B_2}L({f^{**}}) + {B_3}L({f^{n + 1}})} \right) + \frac{1}{2}\Delta {t^2}\left( {{C_0}\frac{\partial }{{\partial t}}L({f^n}) + {C_1}\frac{\partial }{{\partial t}}L({f^*}) + {C_2}\frac{\partial }{{\partial t}}L({f^{**}}) + {C_3}\frac{\partial }{{\partial t}}L({f^{n + 1}})} \right) \\ & + \Delta t\left( {{{ B_0}} \Omega ({f^n}) + {{ B_1}} \Omega ({f^*}) + {{ B_2}} \Omega ({f^{**}}) + {{ B_3}} \Omega ({f^{n + 1}})} \right) + \frac{1}{2}\Delta {t^2}\left( {{{ C_0}} \frac{\partial }{{\partial t}}\Omega ({f^n}) + {{ C_1}} \frac{\partial }{{\partial t}}\Omega ({f^*}) + {{ C_2}} \frac{\partial }{{\partial t}}\Omega ({f^{**}}) + {{ C_3}} \frac{\partial }{{\partial t}}\Omega ({f^{n + 1}})} \right),
\end{split}
\label{EvolutionEquationFirst}
\end{equation}
where $B_0,B_1,B_2,B_3,C_0,C_1,C_2,C_3, A$  and $K$  are parameters which can be adjusted to fulfill requirements in accuracy.

To determine these parameters, the terms of $M (f)$ in Eq. \eqref{EvolutionEquationFirst_M} are expanded about $f^n$,
\begin{equation}
	\small
	\begin{split}
	\Delta t&\left( {{B_0}M({f^n}) + {B_1}M({f^*}) + {B_2}M({f^{**}}) + {B_3}M({f^{n + 1}})} \right) + \frac{{\Delta {t^2}}}{2}\left( {{C_0}\frac{\partial }{{\partial t}}M({f^n}) + {C_1}\frac{\partial }{{\partial t}}M({f^*}) + {C_2}\frac{\partial }{{\partial t}}M({f^{**}}) + {C_3}\frac{\partial }{{\partial t}}M({f^{n + 1}})} \right) \\
	=&\Delta t({B_0} + {B_1} + {B_2} + {B_3})M({f^n})
	 + \frac{{\Delta {t^2}}}{2}\left( {A{B_1} + K{B_2} + {B_3} + \frac{1}{2}({C_0} + {C_1} + {C_2} + {C_3})} \right){M_f}({f^n})M({f^n})\\
	 + &\frac{{\Delta {t^3}}}{6}\left( {3({A^2}{B_1} + {K^2}{B_2} + {B_3} + A{C_1} + K{C_2} + {C_3})} \right) \left( {{M_f}^2({f^n})M({f^n}) + {M_{ff}}({f^n}){M^2}({f^n})} \right) + O(\Delta {t^4}),
	\end{split}
	\label{ExpansionM}
\end{equation}

Comparing Eq. \eqref{ExpansionOri} with Eq. \eqref{ExpansionM}, if the constructed solution Eq. \eqref{EvolutionEquationFirst_M} is required to be in the third-order of accuracy in time, the following relationships can be obtained
\begin{equation}
\small
	B_0+B_1+B_2+B_3=1,AB_1 + KB_2 + B_3+\frac{1}{2}(C_0+C_1+C_2+C_3)=\frac{1}{2},A^2B_1+K^2B_2+B_3+AC_1+KC_2+C_3=\frac{1}{3}.
	\label{Para_All}
\end{equation}
It should be noticed that the number of parameters is more than the number of equations, due to which some parameters can be artificially adjusted.

The DUGKS offers the flexibility in treating transport and collision terms; these two terms are respectively discretized by the midpoint rule and the trapezoidal rules respectively \cite{guo2013discrete}. According to this feature, the parameters of $L(f)$ and $\Omega(f)$ in Eq. \eqref{EvolutionEquationFirst} are revised as
\begin{equation}
\footnotesize
\begin{split}
{f^{n + 1}}{\rm{ }} = {f^n}  & + \Delta t\left( {{{\tilde B_0}} L ({f^n}) + {{\tilde B_1}} L ({f^*}) + {{\tilde B_2}} L ({f^{**}}) + {{\tilde B_3}} L ({f^{n + 1}})} \right) + \frac{1}{2}\Delta {t^2}\left( {{{\tilde C_0}} \frac{\partial }{{\partial t}}L ({f^n}) + {{\tilde C_1}} \frac{\partial }{{\partial t}}L ({f^*}) + {{\tilde C_2}} \frac{\partial }{{\partial t}}L ({f^{**}}) + {{\tilde C_3}} \frac{\partial }{{\partial t}}L ({f^{n + 1}})} \right) \\
 & + \Delta t\left( {{{\overline B_0}} \Omega ({f^n}) + {{\overline B_1}} \Omega ({f^*}) + {{\overline B_2}} \Omega ({f^{**}}) + {{\overline B_3}} \Omega ({f^{n + 1}})} \right) + \frac{1}{2}\Delta {t^2}\left( {{{\overline C_0}} \frac{\partial }{{\partial t}}\Omega ({f^n}) + {{\overline C_1}} \frac{\partial }{{\partial t}}\Omega ({f^*}) + {{\overline C_2}} \frac{\partial }{{\partial t}}\Omega ({f^{**}}) + {{\overline C_3}} \frac{\partial }{{\partial t}}\Omega ({f^{n + 1}})} \right),
\end{split}
\label{EvolutionEquationFirst_L_O}
\end{equation}
Eq. \eqref{EvolutionEquationFirst_L_O} offers a more general form of DUGKS in the third-order of accuracy in time. Therefore, the original DUGKS can be considered as a special case of Eq. \eqref{EvolutionEquationFirst_L_O} by choosing a specific set of parameters as shown in Remark I. 

Similar comparison with Eq. \eqref{ExpansionOri} can be carried out, which gives the following system of equations that determine the parameter of $L(f)$ and $\Omega(f)$ should also be the solutions of Eq. \eqref{Para_All}:
\begin{subequations}
	\small
	\begin{equation}
	{\tilde B}_0+{\tilde B}_1+{\tilde B}_2+{\tilde B}_3=1,A{\tilde B}_1 + K{\tilde B}_2 + {\tilde B}_3+\frac{1}{2}({\tilde C}_0+{\tilde C}_1+{\tilde C}_2+{\tilde C}_3)=\frac{1}{2},A^2{\tilde B}_1+K^2{\tilde B}_2+{\tilde B}_3+A{\tilde C}_1+K{\tilde C}_2+{\tilde C}_3=\frac{1}{3}.
	\label{Para_L}
	\end{equation}
	\begin{equation}
	{\overline B}_0+{\overline B}_1+{\overline B}_2+{\overline B}_3=1,A{\overline B}_1 + K{\overline B}_2 + {\overline B}_3+\frac{1}{2}({\overline C}_0+{\overline C}_1+{\overline C}_2+{\overline C}_3)=\frac{1}{2},A^2{\overline B}_1+K^2{\overline B}_2+{\overline B}_3+A{\overline C}_1+K{\overline C}_2+{\overline C}_3=\frac{1}{3}.
	\label{Para_O}
	\end{equation}
	\label{Para_L_O}
\end{subequations}

\begin{remark}[I]
	\label{remark1}
	If the second-order of temporal accuracy is required, parameter of the third-order terms are omitted. And the remaining parameters are determined by
	\begin{equation*}
	{\tilde B}_0+{\tilde B}_1+{\tilde B}_2+{\tilde B}_3=1,A{\tilde B}_1 + K{\tilde B}_2 + {\tilde B}_3+\frac{1}{2}({\tilde C}_0+{\tilde C}_1+{\tilde C}_2+{\tilde C}_3)=\frac{1}{2},
	\end{equation*}
	\begin{equation*}
	{\overline B}_0+{\overline B}_1+{\overline B}_2+{\overline B}_3=1,A{\overline B}_1 + K{\overline B}_2 + {\overline B}_3+\frac{1}{2}({\overline C}_0+{\overline C}_1+{\overline C}_2+{\overline C}_3)=\frac{1}{2}.
	\end{equation*}
	
	The original second-order time-accurate DUGKS can be obtained by setting the parametric values as
	\begin{equation*}
	\begin{array}{c}
	A=0, K=1/2, {\tilde B}_0 = 0, {\tilde B}_1 = 0, {\tilde B}_2=1, {\tilde B}_3=0, {\tilde C}_0={\tilde C}_1={\tilde C}_2={\tilde C}_3=0, \\
	{\overline B}_0 = 1/2, {\overline B}_1 ={\overline B}_2= 0, {\overline B}_3=1/2, {\overline C}_0={\overline C}_1={\overline C}_2={\overline C}_3=0.
	\end{array}
	\end{equation*}
	
	The evolution equation is then given as
	\begin{equation}
	\small
	\begin{split}
	f^{n+1} = f^n + \frac{1}{2} \Delta t L(f^*) + \frac{1}{2} \Delta t \left( \Omega(f^n)  +  \Omega(f^{n+1}) \right) + O(\Delta t^3).
	\end{split}
	\label{feqIntD2nd}
	\end{equation}

The explicit treatment of Eq. \eqref{feqIntD2nd} yields
\begin{equation}
\tilde f^{n+1} = \tilde f^{+,n} + \frac{1}{2} \Delta t L(f^*) + O(\Delta t^3).
\label{feqIntD2nd_Explicit}
\end{equation}
where
\begin{equation*}
\tilde f = f - {1 \over 2}\Delta t \Omega,
\end{equation*}
\begin{equation*}
\tilde f^+ = \frac{2 \tau - \Delta t}{2 \tau + \Delta t} \tilde f + \frac{2 \Delta t}{2 \tau + \Delta t} f^{eq}.
\end{equation*}
And Eq. \eqref{feqIntD2nd_Explicit} is the practical evolution equation of the original DUGKS. Clearly, the original DUGKS is a 2nd-order method.
	
	%The error term of $L(f)$ is $O(\xi^2h^2)$, where $h = \Delta t / 2$. Meanwhile, $\xi dt = a \Delta x$, also $O(\Delta x^2)$

\end{remark}

As mentioned above, the number of parameters is more than the number of equations. Therefore, some parameters can be artificially determined without sacrificing the overall order of accuracy. Firstly, to use the DUGKS flux reconstruction scheme which is designed to solve the flux at the intermediate time steps, we want to keep $L(f^*)$ and $L(f^{**})$ only. Thus, some additional conditions are added to the indeterminate equations \eqref{Para_L_O}
\begin{equation*}
	{\tilde B}_0 = 0,{\tilde B}_3 = 0,{\tilde C}_0={\tilde C}_1={\tilde C}_2={\tilde C}_3=0.
\end{equation*}
Solutions to Eq. \eqref{Para_L} can then be written as functions of $K$
\begin{equation*}
	A=\frac{3K-2}{6K-3}, {\tilde B}_0 = 0, {\tilde B}_1 = \frac{3(2K-1)^2}{4(3K^2-3K+1)}, {\tilde B}_2=\frac{1}{4(3K^2-3K+1)}, {\tilde C}_0={\tilde C}_1={\tilde C}_2={\tilde C}_3=0,
\end{equation*}
Note that $A < K$, and all the parameters should be kept within $[0,1]$, which gives the range of $K$ as
\begin{equation*}
	\frac{2}{3} < K < 1 .
\end{equation*}

%Then we set $K=\frac{3}{4}$, and the solution is
%\begin{equation*}
%A=1/6, B_0 = 0, B_1 = 3/7, B_2=4/7, C_0=C_1=C_2=C_3=0,
%\end{equation*}

The second consideration is the explicitness of the scheme. Under this consideration, the collision terms $\Omega(f^*)$ and $\frac{\partial}{\partial t} \Omega(f^*)$ should be omitted, which gives
\begin{equation*}
	{\overline B}_1={\overline B}_2= 0,{\overline C}_1 = {\overline C}_2 =0.
\end{equation*}
All parameters in equations \eqref{Para_O} can now be expressed as functions of ${\overline B}_3$
\begin{equation*}
	{\overline B}_0 = 1 - {\overline B}_3,{\overline B}_1 = 0,{\overline B}_2 = 0,{\overline C}_0 = 2/3-{\overline B}_3,{\overline C}_1 = 0,{\overline C}_2 = 0,{\overline C}_3 = 1/3-{\overline B}_3.
\end{equation*}
It should be noticed that $A$ and $K$ are automatically eliminated in the above solutions.

Thus, the evolution equation becomes
\begin{equation}
{f^{n + 1}}{\rm{ }} = {f^n}+ \frac{{\Delta t}}{{4(3{K^2} - 3K + 1)}}\left( {3{{(2K - 1)}^2}L({f^*}) + L({f^{**}})} \right) + \frac{1}{2}\Delta {t^2}\left( {({\frac{2}{3}-{\overline B}_3}) \frac{\partial }{{\partial t}}\Omega ({f^n}) + ({\frac{1}{3}-{\overline B}_3}) \frac{\partial }{{\partial t}}\Omega ({f^{n + 1}})} \right).
\label{EvolutionEquationPara}
\end{equation}

Determining a specific value of ${\overline B}_3$ is mathematically unnecessary. The second-order discretization of the temporal gradient term $\frac{\partial }{{\partial t}}\Omega ({f})$ in Eq. \eqref{EvolutionEquationPara} automatically vanishes ${\overline B}_3$:
\begin{subequations}
	\begin{equation}
	\frac{\partial }{{\partial t}}\Omega ({f^n}) = \frac{{(1 - {a^2})\Omega ({f^n}) - \Omega ({f^{'}}) + {a^2}\Omega ({f^{n + 1}})}}{{a(a - 1)\Delta t}} + O(\Delta {t^2}),
	\end{equation}
	
	\begin{equation}
	\frac{\partial }{{\partial t}}\Omega ({f^{n + 1}}) = \frac{{ - (1 - {a^2})\Omega ({f^n}) + \Omega ({f^{'}}) - a(2 - a)\Omega ({f^{n + 1}})}}{{a(a - 1)\Delta t}} + O(\Delta {t^2}),
	\end{equation}
\end{subequations}
where $0 < a < 1$ and $f^{'}=f(x,\xi,t+a\Delta t)$. And the evolution equation Eq. \eqref{EvolutionEquationPara} becomes
\begin{equation}
\begin{split}
{f^{n + 1}} = {f^n} &+ \frac{{\Delta t}}{{4(3{K^2} - 3K + 1)}}\left( {3{{(2K - 1)}^2}L({f^*}) + L({f^{**}})} \right)  \\ & + \Delta t\left( P_n\Omega ({f^n}) + P^{'}\Omega ({f^{'}}) + P_{n+1}\Omega ({f^{n + 1}}) \right)  + O(\Delta {t^4}),
\label{feqIntD3rd2}
\end{split}
\end{equation}
where ${P_n} = \frac{{1 - 4a + 3{a^2}}}{{6a(a - 1)}}$, ${P^{'}} = -\frac{1}{{6a(a - 1)}}$ and ${P_{n + 1}} = \frac{{ - 2a + 3{a^2}}}{{6a(a - 1)}}$ are the coefficients of $\Omega (t_n)$, $\Omega ({f^{'}})$ and $\Omega (t_{n+1})$. Note that ${P_n} + {P^{'}} + {P_{n + 1}} = 1$. 

\begin{figure}
	\centering
	\includegraphics[width=3.2in,height=3in]{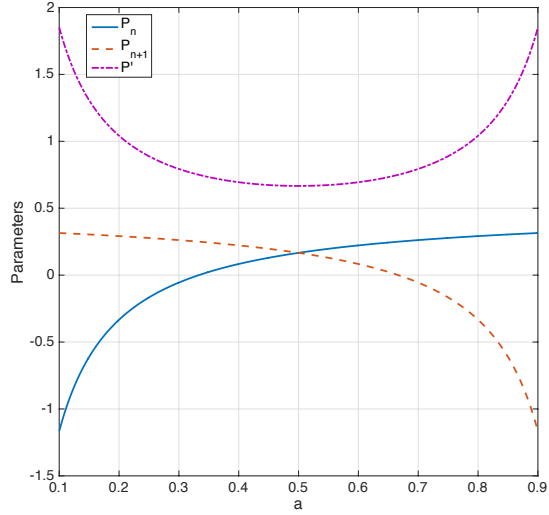}
	\caption{Parameters $P_n, P^{'}$ and $P_{n+1}$ with $a$.} 
	\label{fig:Parameters_P}
\end{figure}

Two free parameters $a$ and $K$, still exist in the evolution equation. To determine their values, we firstly analyse the range of parameters ${P_n}, {P^{'}}$ and $ {P_{n + 1}}$. Fig. \ref{fig:Parameters_P} shows the variations of ${P_n}, {P^{'}}$ and $ {P_{n + 1}}$ with $a$.
We also set $0 \le P_n, a, P_{n+1} \le 1$ to ensure that the present method is an interpolation scheme in time. And the range of $a$ will determined as
\begin{equation*}
\frac{1}{3} \le a \le \frac{1}{2}.
\end{equation*}

To perform explicit treatment of Eq. \eqref{feqIntD3rd2}, we set
\begin{subequations}
	\begin{equation}
	\hat f = f - P_{n+1}\Delta t \Omega,
	\end{equation}
	
	\begin{equation}
	{\hat f^ + } = \hat f + ({P_n} + {P_{n + 1}})\Delta t\Omega  = \frac{{\tau  - {P_n}\Delta t}}{{\tau  + {P_{n + 1}}\Delta t}}\hat f + \frac{{({P_n} + {P_{n + 1}})\Delta t}}{{\tau  + {P_{n + 1}}\Delta t}}{f^{eq}}.
	\end{equation}
\end{subequations}

Then the evolution equation becomes
\begin{equation}
{\hat f^{n + 1}} = {\hat f^{+,n}} + \Delta t P^{'}\Omega ({f^{'}})  + \frac{{\Delta t}}{{4(3{K^2} - 3K + 1)}}\left( {3{{(2K - 1)}^2}L({f^*}) + L({f^{**}})} \right)  + O(\Delta {t^4}).
\label{EEWithP}
\end{equation}
In above equation, $\hat f$ will be updated rather than original $f$. 

The right-hand side of Eq. \eqref{EEWithP} can be expanded about $\hat f^n$, and the overall parameter ${\hat P_n}$ of $\hat f^n$ is assessed from the perspective of numerical stability. A simple Von Neumann analysis is employed by neglecting the effect of term $L({f})$ on the interface points at the intermediate time-steps, which leads to the following stable region of ${\hat P_n}$
\begin{equation*}
{\hat P_n} = \left| {\frac{{ - 3{a^2}{D^2} + 13a{D^2} + 10aD - 22D + 32}}{{4(aD + 2)(D + 4)}}} \right| \le 1,
\end{equation*}
where $D = \Delta t / \tau$.

\begin{figure}
	\centering
	\includegraphics[width=3.2in,height=3in]{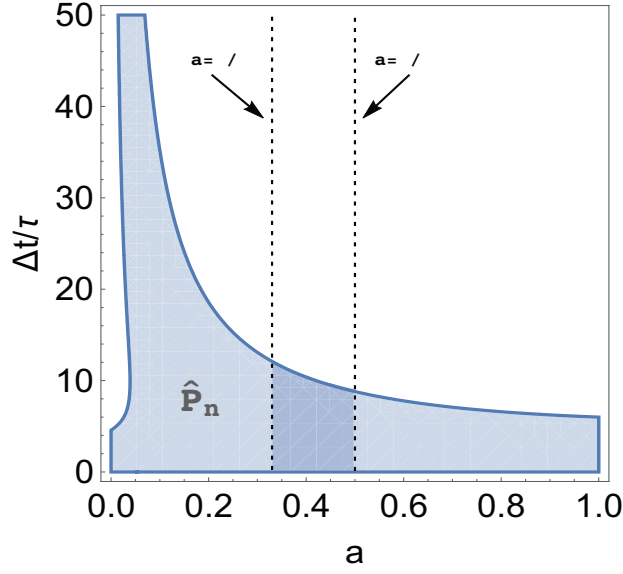}
	\caption{Available region for $\hat P_n$ with $a$ and $\Delta t / \tau$.} 
	\label{fig:Parameter_a}
\end{figure}

Fig. \ref{fig:Parameter_a} shows the stability region of $\hat P_n$. The maximum stability region is obtained when setting $a = 1/3$ and $\Delta t / \tau \le 12$. Apparently, the stability region of the present scheme is larger than traditional LBE based kinetic methods which only allow the tolerance of $\Delta t / \tau \le 2$ \cite{guo2005finite, wang2007implicit}. Numerical validation of the above analysis will be presented in Sec. \ref{sec:LDF}.

Noted that the term of $\Omega(f')$ in Eq. \eqref{EEWithP} at $t' = t_n + a\Delta t$ can be evaluated by the original T2S2-DUGKS method \cite{guo2013discrete}, and the truncated error of $\Delta t \Omega(f')$ is in the order of $O(\Delta t^4)$ which is consisted with the overall order of accuracy of the presented method. Furthermore, if we set $A=a/2$, the term of $L(f^*)$ at $t_* = t_n + A\Delta t$ in Eq. \eqref{EEWithP} can be directly used to compute $\Omega(f')$, which will greatly lower the computational efforts.

Under the above considerations, we set $a=1/3$, then $A = 1/6$ and $K=3/4$ can be calculated from the equations: $A=a/2$ and $A=\frac{3K-2}{6K-3}$. The evolution equation is then reduced to
\begin{equation}
	{\hat f}^{n+1} = {{\hat f}^{+,n} } + \frac{3}{4} \Delta t \Omega(f')  + \frac{1}{7} \Delta t \left(3 L(f^*) + 4 L(f^{**}) \right),
	\label{feqIntD3rdHat2}
\end{equation}
where
\begin{equation}
	\hat f = f - {1 \over 4}\Delta t \Omega, {{\hat f}^ + } = \hat f + {1 \over 4}\Delta t \Omega ,
\end{equation}
and the intermediate time steps are $t_* = t_n + \Delta t / 6$, $t' = t_n + \Delta t / 3$ and  $t_{**} = t_n + 3\Delta t / 4$.

Some useful equations can be derived for computation:
\begin{equation}
	\hat f^+ = \frac{4 \tau}{4 \tau + \Delta t} \hat f + \frac{ \Delta t}{4 \tau + \Delta t} f^{eq},
\end{equation}
\begin{equation}
\tilde f^+ = \frac{6 \tau - \Delta t}{6 \tau + \Delta t} \tilde f + \frac{2 \Delta t}{6 \tau + \Delta t} f^{eq},
\label{ftp}
\end{equation}
%\begin{equation}
%	{\bar f^{+,b \Delta t}} = \frac{4\tau - 2b\Delta t}{4\tau + \Delta t}\hat f + \frac{(2b + 1)\Delta t}{4\tau + \Delta t} f^{eq}
%\end{equation}
\begin{equation}
	\Delta t \Omega(f') = -\frac{6 \Delta t}{6 \tau + \Delta t} (\tilde{f} - f^{eq}),
	\label{collisionterm}
\end{equation}
where $\tilde f = f - {1 \over 6}\Delta t \Omega$ equals to the definition of $\tilde f$ in T2S2-DUGKS with the time-step $\Delta t / 3$.

\subsubsection{Space discretization}
In this section, the evolution equation Eq. \eqref{feqIntD3rdHat2} will be discretized in space by finite-volume method (FVM). The flow domain is divided into a set of control volumes, and each control volume $V_j$ is centered at $\bm{x}_j$. The discrete form of evolution equation at cell center is then given by
\begin{equation}
{\hat f_j}^{n+1} = {{\hat f_j}^{+,n} } + \frac{3}{4} \Delta t \Omega'_j  + \frac{1}{7} \Delta t \left(3 {L}^*_j + 4 {L}^{**}_j \right),
\label{feqIntD3rdHat2_DS}
\end{equation}
where the cell-averaged values, such as ${\hat f_j}^{n}$ and $\Omega'_j$, are evaluated from the discretized distribution function
\begin{equation}
f_j^n = \frac{1}{{\left| {{V_j}} \right|}}\int_{{V_j}} {f(\bm{x},\bm{\xi} ,{t_n})} {\kern 1pt} d\bm{x},
\end{equation}
and the micro-fluxes $L^*_j = L_j(t_*)$ and $L^{**}_j = L_j(t_{**})$ across the cell interface at time $t_* = t_n+\Delta t / 6$ and $t_{**} = t_n+3\Delta t/4$ are given by
\begin{equation}
\label{microflux}
{L_j(t)} = \frac{1}{{\left| {{V_j}} \right|}} \int_{\partial {V_j}} {(\bm{\xi}  \cdot \bm{n})f(\bm{x},\bm{\xi} ,{t})} {\kern 1pt} d\bm{x},
\end{equation}

The key point in calculating micro-fluxes is to estimate the distribution function $f$ on the cell interface $x_b$ at the time $t_*$ and $t_{**}$. The DUGKS reconstruction process \cite{guo2013discrete} can be used here. However, higher order of accuracy is required. To construct the high-order scheme, we start from Boltzmann equation,
\begin{equation}
	{{\partial f} \over {\partial t}} + \xi  \cdot \nabla f = {{df(x + \xi t,\xi ,t)} \over {dt}} = \Omega. 
\end{equation}
Integrate it along the characteristic line $x + \xi t$ over $[0,h]$, 
\begin{equation}
	f(x + \xi t,\xi ,t) - f(x,\xi ,t) = \int_0^{h} {\Omega (x + \xi s,\xi ,t + s)} ds,
	\label{flux_BE_Char}
\end{equation}
where $h$ is the time step depending on the DUGKS reconstruction scheme. In the presented method， it should be $h = A \Delta t = \Delta t/6$ or $h = (K-2A)\Delta t = 5\Delta t/12$.

Adopting the trapezoidal rule to approximate the integration in Eq. \eqref{flux_BE_Char} gives
%\begin{equation}
%	f(x + \xi h,\xi ,t+h) - f(x,\xi ,t) = {{h} \over 2}\left( {\Omega (x + \xi h,\xi ,t + h) + \Omega (x,\xi ,t)} \right) + {\rm O}({h^3})
%\end{equation}
%or
\begin{equation}
\label{DF_INTER}
	f(x,\xi ,t + h) - f(x - \xi h,\xi ,t) = {h \over 2}\left( {\Omega (x,\xi ,t + h) + \Omega (x - \xi h,\xi ,t)} \right).
\end{equation}
Similar to the treatment in Eq. \eqref{EEWithP}, a new variable $\bar f$ can be defined to remove the implicity in Eq. \eqref{DF_INTER},
\begin{equation}
\label{original_f}
\bar f = f - \frac{{h}}{2}\Omega, 
\end{equation}
or
\begin{equation}
f = \frac{{2\tau }}{{2\tau  + h}}\bar f + \frac{h}{{2\tau  + h}}{f^{eq}}.
\label{recover2f}
\end{equation}

Then Eq. \eqref{DF_INTER} can also be rewritten as
\begin{equation}{\label{interfaceEvof}}
	\bar f({\bm{x}_b},\bm{\xi} ,{t} + h) = {\bar f^{+,h} }({\bm{x}_b} - \bm{\xi} h,\bm{\xi} ,{t}),
\end{equation}
where
\begin{subequations}

	\begin{equation}
		{\bar f^{+,h} } = \frac{{2\tau  - h}}{{2\tau  + h}}\bar f + \frac{{2h}}{{2\tau  + h}}{f^{eq}}.
	\end{equation}
\end{subequations}

The term ${\bar f^ + }({\bm{x}_b} - \bm{\xi} h,\bm{\xi} ,{t})$ can be approximated from the Taylor-series expansion analysis with desired order of accuracy,
\begin{equation}
	\small
	{\bar f^ + }({x_b} - {\rm{ }}\xi h) = \left\{ {\begin{array}{*{20}{l}}
			{{{\bar f}^ + }({x_j}) + ({x_b} - {\rm{ }}\xi h - {x_j}) \cdot \nabla {{\bar f}^ + }({x_j}) + {{({x_b} - {\rm{ }}\xi h - {x_j})}^2}:{\nabla ^2}{{\bar f}^ + }({x_j}),}\quad &in \quad cell \quad V_j\\
			{{{\bar f}^ + }({x_{j + 1}}) + ({x_b} - {\rm{ }}\xi h - {x_{j + 1}}) \cdot \nabla {{\bar f}^ + }({x_{j + 1}}) + {{({x_b} - {\rm{ }}\xi h - {x_{j + 1}})}^2}:{\nabla ^2}{{\bar f}^ + }({x_{j + 1}})},\quad &in \quad cell \quad V_{j+1}
	\end{array}} \right.
\label{flux_DS}
\end{equation}
Space derivatives $\nabla {{\bar f}^ + }$ and ${\nabla ^2}{{\bar f}^ + }$ can be calculated by numerical methods, such as the midpoint scheme.

After substituting Eq. \eqref{flux_DS}, \eqref{interfaceEvof} and  \eqref{recover2f} into Eq. \eqref{microflux}, the flux $L_j$ can be determined.
%the total error term of $L_j$ in space should be $O(\xi^3h^3\Delta x)$. Meanwhile, $\xi \Delta t = a \Delta x$, also $O(\Delta x^4)$. 
%Thus, the error term of the evolution equation is
%\begin{equation}
%{\hat f_j}^{n+1} = {{\hat f_j}^{+,n} } + \frac{3}{4} \Delta t \Omega'_j  + \frac{1}{7} \Delta t \left(3 {L}^*_j + 4 {L}^{**}_j \right)  + O(\Delta t^4 + \Delta t \Delta x^4).
%	\label{feqIntD3rdHat2WithErroTerm}
%\end{equation}
Additionally, treating Eq. \eqref{microflux} with the midpoint integral formula seems to influence the order of accuracy in space. But it is easier in application and faster in computation. And there is no significant effect shown from the numerical results. So, we still use the midpoint integral formula in present method.

%With the definition of CFL condition, we can know that $\Delta t$ and $\Delta x / \xi$ are in the same order. Hence, the presented scheme can be seen as a third-order scheme in both time and space.

\begin{remark}[II]
    The flux reconstruction process in DUGKS is also the major constrain in achieving higher order of temporal accuracy. It can be shown in the trapezoidal rule adopted in space discretization
	\begin{equation}
	f(x + \xi h,\xi ,t+h) - f(x,\xi ,t) = {{h} \over 2}\left[ {\Omega (x + \xi h,\xi ,t + h) + \Omega (x,\xi ,t)} \right] + {\rm O}({h^3}),
	\end{equation}
	the $O(\Delta t^3)$ will be remained in the term of $L_j$, and $O(\Delta t^4)$ for the term of $\Delta t L_j$ in time evolution. Thus, forth or even higher order DUGKS scheme cannot be constructed before the flux reconstruction method is revised to higher-order of accuracy.
\end{remark}

Useful equations:
\begin{equation}
{\bar f^{+,\frac{1}{6} \Delta t} }(\bm{x},\bm{\xi},t_n) = \frac{{12\tau  - \Delta t}}{{12\tau  + 3\Delta t}}\hat f(\bm{x},\bm{\xi},t_n) + \frac{{4\Delta t}}{{12\tau  + 3\Delta t}}{f^{eq}}(\bm{x},\bm{\xi},t_n),
\label{f_bar_1_6}
\end{equation}
\begin{equation}
\tilde{f}(\bm{x},\bm{\xi},t_n)=\frac{5}{4}{\bar f^{+,\frac{1}{6} \Delta t} }(\bm{x},\bm{\xi},t_n) - \frac{1}{4}\hat{f}(\bm{x},\bm{\xi},t_n),
\label{f_tilde_1_3}
\end{equation}
\begin{equation}
{\bar f^{+,\frac{5}{12} \Delta t} }(\bm{x},\bm{\xi},t_n + \frac{1}{3} \Delta t) = \frac{{24\tau  - 5\Delta t}}{{24\tau  + 10\Delta t}}\tilde f(\bm{x},\bm{\xi},t_n + \frac{1}{3} \Delta t) + \frac{{15\Delta t}}{{24\tau  + 10\Delta t}}{f^{eq}}(\bm{x},\bm{\xi},t_n + \frac{1}{3} \Delta t),
\label{fbp_5_12}
\end{equation}
\begin{equation}
\hat{f}^+(\bm{x},\bm{\xi},t_n)=\frac{1}{4}{\hat f}(\bm{x},\bm{\xi},t_n) + \frac{3}{4}{\bar f^{+,\frac{1}{6} \Delta t} }(\bm{x},\bm{\xi},t_n).
\label{fhp}
\end{equation}

\subsubsection{Computational sequence of the third-order DUGKS}
\label{sec:algorithm_of_T3S3}
\begin{figure}
	\centering
	\includegraphics[width=6.5in,height=2.5in]{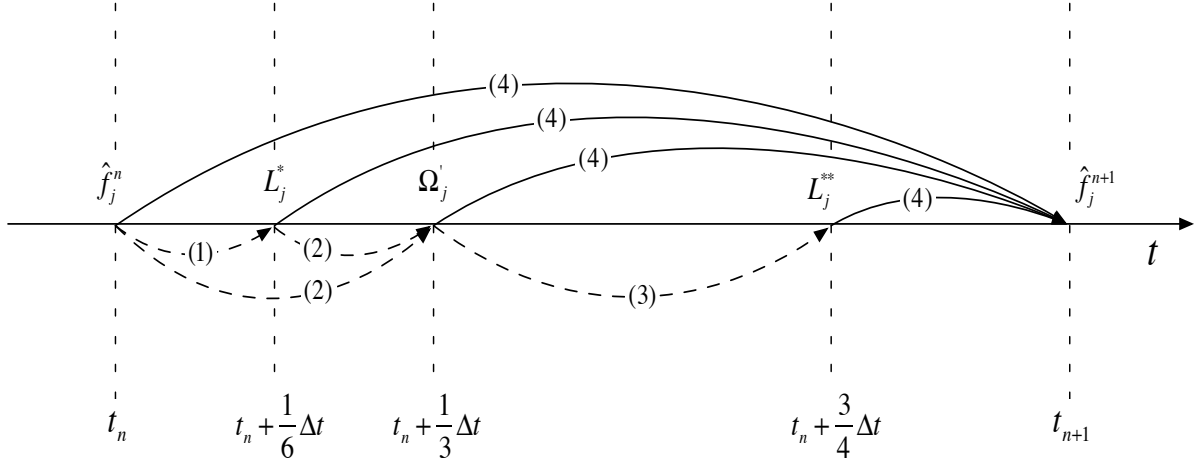}
	\caption{Time-line of T3S3-DUGKS.} 
	\label{fig:T3S3_Timeline}
\end{figure}

From the evolution equation Eq. \eqref{feqIntD3rdHat2}, it can be found that the key point to evolve $\hat f$ is to calculate the intermediate properties ${L}^*_j$, $\Delta t \Omega(\bm{x_j}, \bm{\xi},t')$, ${L}^{**}_j$ at the intermediate times. Fig. \ref{fig:T3S3_Timeline} illustrates the time-line of the evolution process and details are shows as follows,

\paragraph{Step (1)}
Use the DUGKS reconstruct scheme to calculate ${L}^*_j$ from $\hat f(\bm{x_j},\bm{\xi},t_n)$.
\begin{equation*}
\hat f(\bm{x_j},\bm{\xi},t_n) \xlongrightarrow{\eqref{f_bar_1_6}}{} {\bar f^{+,\frac{1}{6} \Delta t} }(\bm{x_j},\bm{\xi},t_n) \xlongrightarrow{\eqref{interfaceEvof} \eqref{flux_DS}}{} \bar f({\bm{x}_b},\bm{\xi} ,{t_*}) \xlongrightarrow{\eqref{recover2f}}{} f({\bm{x}_b},\bm{\xi} ,{t_*}) \xlongrightarrow{\eqref{microflux}}{} {L}_j(t_*)
\end{equation*}

\paragraph{Step (2)}
Estimate $\hat f(\bm{x_j},\bm{\xi},t')$ with the T2S2-DUGKS method, and compute $\Delta t \Omega(\bm{x_j},\bm{\xi},t')$.
\begin{equation*}
\hat f(\bm{x_j},\bm{\xi},t_n) \xlongrightarrow{{\bar f^{+,\frac{1}{6} \Delta t} }(\bm{x_j},\bm{\xi},t_n) , \eqref{f_tilde_1_3}}{} \tilde{f}(\bm{x_j},\bm{\xi},t_n) \xlongrightarrow{\eqref{ftp}}{} \tilde{f}^+(\bm{x_j},\bm{\xi},t_n) \xlongrightarrow{{L}_j(t_*), \eqref{feqIntD2nd_Explicit}}{} \tilde{f}(\bm{x_j},\bm{\xi},t') \xlongrightarrow{\eqref{collisionterm}}{} \Delta t \Omega(\bm{x_j},\bm{\xi},t')
\end{equation*}

Noted that the time step in Eq. \eqref{feqIntD2nd_Explicit} should be $\Delta t / 3$ instead.

\paragraph{Step (3)}
Use the DUGKS reconstruct scheme again to calculate ${L}^{**}_j$ from $\hat f(\bm{x_j},\bm{\xi},t')$.
\begin{equation*}
\tilde{f}(\bm{x_j},\bm{\xi},t') \xlongrightarrow{\eqref{fbp_5_12}}{}{\bar f^{+,\frac{5}{12} \Delta t} }(\bm{x_j},\bm{\xi},t') \xlongrightarrow{\eqref{interfaceEvof} \eqref{flux_DS}}{} \bar f({\bm{x}_b},\bm{\xi} ,{t_{**}}) \xlongrightarrow{\eqref{recover2f}}{} f({\bm{x}_b},\bm{\xi} ,{t_{**}}) \xlongrightarrow{\eqref{microflux}}{} {L}_j(t_{**})
\end{equation*}

\paragraph{Step (4)}
Calculate $\hat f(\bm{x_j},\bm{\xi},t_{n+1})$ from the obtained terms of $\hat f(\bm{x_j},\bm{\xi},t_n), {L}^*_j, \Delta t \Omega(\bm{x_j},\bm{\xi},t'), {L}^{**}_j$ and ${L}^{**}_j$.
\begin{equation*}
\hat f(\bm{x_j},\bm{\xi},t_n) \xlongrightarrow{{\bar f^{+,\frac{1}{6} \Delta t} }(\bm{x_j},\bm{\xi},t_n) , \eqref{fhp}}{} \hat{f^+}(\bm{x_j},\bm{\xi},t_n) \xlongrightarrow{{L}_j(t_*), \Delta t \Omega(\bm{x_j},\bm{\xi},t'), {L}_j(t_{**}), \eqref{feqIntD3rdHat2_DS}}{} \hat{f}(\bm{x_j},\bm{\xi},t_{n+1}) 
\end{equation*}
Actually, the memory cost in computer program of DUGKS-T3S3 is only up to $\frac{4}{3}$ times more than DUGKS-T2S2. It will be shown in Section \ref{sec:micro_cavity}.

\subsection{Discrete particle velocities}
\label{sec:discrete_velocities}
The discrete particle velocities and associated weights can be obtained by the quadrature rules \cite{guo2013discrete} in one dimension. By using the tensor product method, we can also get the discretized particle velocity space in higher dimensions \cite{shan2006kinetic, la2015multispeed}. For continuum flow, the three-point Gauss-Hermite quadrature are employed to obtain the one-dimensional discrete particle velocities and associated weights,
\begin{equation*}
	{\bm{\xi}} = \sqrt {3RT} \left[ {\begin{array}{*{20}{c}}
			{ - 1}& \ 0& \ 1
	\end{array}} \right], \ {W} = \left[ {\begin{array}{*{20}{c}}
	 \frac{1}{3}& \ \frac{2}{3}& \ \frac{1}{3},
	\end{array}}\right]
\end{equation*}
and the discrete velocities and associated weights in 2D space can be derived as
\begin{equation*}
	{\bm{\xi}} =  \sqrt {3RT}\left[ {\begin{array}{*{20}{c}}
			{ - 1}&{ - 1}&{ - 1}& \ 0& \ 0& \ 0& \ 1& \ 1& \ 1\\
			{ - 1}& \ 0& \ 1&{ - 1}& \ 0& \ 1&{ - 1}& \ 0& \ 1
	\end{array}} \right] , 
\end{equation*}
\begin{equation*}
\qquad {W} = \left[\begin{array}{*{20}{c}}
\frac{1}{{36}}& \ \frac{1}{9}& \ \frac{1}{{36}}& \ \frac{1}{9}& \ \frac{4}{9}& \ \frac{1}{9}& \ \frac{1}{{36}}& \ \frac{1}{9} & \ \frac{1}{{36}}
\end{array}
 \right] 
\end{equation*}
For rarefied flow, the particle velocity space will be discretized by the Newton-Cotes quadrature with $101 \times 101$ nodes distributed uniformly in the area of $[-4\sqrt{RT},4\sqrt{RT}] \times [-4\sqrt{RT},4\sqrt{RT}]$.

Consequently, the density and the velocity can be calculated discretely by
\begin{equation}
\rho  = \sum\limits_i {{f_i}}, \ {\rho}\bm{u} = \sum\limits_i {{\bm{\xi} _i}{f_i}}.
\end{equation}

%The schematics of the discrete velocities in 2D and 3D spaces can be found in Fig. \ref{fig:DiscreteVelocity_2D}.
%
%
%\begin{figure}
%	\centering
%	\includegraphics[width=2.7in,height=2.7in]{2D_velocity}
%	\caption{The discrete velocities of two-dimensional DUGKS.} 
%	\label{fig:DiscreteVelocity_2D}
%\end{figure}

\subsection{Boundary conditions}
The bounce back (BB) scheme is a commonly used boundary condition in DUGKS, it assumes that velocity of the particle reverses the direction and maintains the absolute value when hitting the wall \cite{ladd1994numerical}. For those particles leaving the wall which is located at a cell interface $\bm{x}_w$, their distribution functions can be determined by
\begin{equation}
\label{BB}
\begin{split}
f({{\bm{x}}_w},{{\bm{\xi }}_i},t + h)& = f({{\bm{x}}_w}, - {{\bm{\xi }}_i},t + h) + 2{\rho _w}{W_i}\frac{{{{\bm{\xi }}_i} \cdot {{\bm{u}}_w}}}{{RT}},\\
&{{\bm{\xi }}_i} \cdot {\bm{n}} < 0,         
\end{split}
\end{equation}
where $\bm{n}$ is the outward unit vector normal to the wall, $\bm{u}_w$ is the wall velocity, and $\rho_w$ is the density of the wall which can be approximated by fluid density $\rho$.

The diffuse-scattering (DS) scheme is used in rarefied flows \cite{guo2013discrete}, which assumes that the distribution function is Maxwellian when the particles reflect from the wall. The DS scheme can be mathematically expressed as
\begin{equation}
f({x_w},{\xi _i},t + h) = {f^{eq}}({\xi _i};{\rho _w},{u_w}),
\end{equation}
where the density $\rho_w$ is obtained by
\begin{equation}
{\rho _w} =  - {\left[ {\sum\limits_{{\xi _i} \cdot n < 0} {({\xi _i} \cdot n){f^{eq}}({\xi _i};1,{u_w})} } \right]^{ - 1}} \times \sum\limits_{{\xi _i} \cdot n > 0} {({\xi _i} \cdot n){f^{eq}}({x_w},{\xi _i},t + h)}. 
\end{equation}

\section{Numerical tests}
\label{sec:numerical_cases}
In this section, four numerical cases, including Taylor vortex flow, doubly periodic shear layers flow, continuum lid-driven cavity flow and micro-cavity flow, will be shown. The first three cases are in continuum regime, and the last one is in rarefied regime. In this section, the mark $TmSn$ represents the method with the $m$-th order of temporal accuracy method and the $n$-th order of spatial accuracy. Through comparison with analytical solutions or benchmark data, numerical accuracy and stability of the present DUGKS-T3S3 method are comprehensively validated.

In our simulations, the initialization of the distribution function is given by its equilibrium part, and the time step $\Delta t$ is determined by the Courant-Friedrichs-Lewy (CFL) condition,
\begin{equation}
	\Delta t = a \frac{{\Delta x}}{C},
\end{equation}
where $a$ is the CFL number, $\Delta x$ is the minimum grid spacing, and $C$ is the maximum discrete velocity. Besides, the Mach and Reynolds numbers appeared in the following simulations are defined by $Ma = U / C$ and $Re = UL/\nu$ with $L$ representing the characteristic length. 

For rarefied flow, the Knudsen number is defined as $Kn = \lambda / L$, where $L$ is the characteristic length of the flow and $\lambda$ is the mean free path of molecular which can be written as
\begin{equation}
\lambda  = \tau \sqrt {\frac{{\pi RT}}{2}}.
\end{equation}

\subsection{Taylor Vortex Flow} The two dimensional incompressible Taylor Vortex Flow is a widely used test case with periodic computation domain. And its analytical solution is
\begin{subequations}
	\begin{equation}
u(x,y,t) =  - \frac{{{u_0}}}{A}\cos (Ax)\sin (By){e^{ - \nu \alpha t}},
	\end{equation}
	\begin{equation}
v(x,y,t) = \frac{{{u_0}}}{B}\sin (Ax)\cos (By){e^{ - \nu \alpha t}},
	\end{equation}
	\begin{equation}
	p(x,y,t) =  - \frac{{u_0^2}}{4}\left[\frac{{\cos (2Ax)}}{{{A^2}}} + \frac{{\cos (2By)}}{{{B^2}}}\right]{e^{ - 2\nu \alpha t}},
	\end{equation}
	\label{TV_AS}
\end{subequations}
where $\bm{u}=(u,v)$ is the fluid velocity, $p$ is the pressure, $\nu$ is the viscosity, $u_0$ is a constant and $\alpha = A^2 + B^2$.

In the simulations, the computation domain is set as $[0, 1] \times [0, 1]$, and $A=B=2\pi$, $u_0=1$, $\nu = 0.01$, $\Delta t = 1 \times 10^{-5}$ and $\Delta t / \tau = 4$. For this decaying flow, the initial distribution function should be calculated by the Navier-Stokes Chapman-Enskog expansion
\begin{equation*}
f(x,\xi ,0) = {f^{eq}} - \tau ({\partial _t}{f^{eq}} + \xi  \cdot \nabla {f^{eq}}),
\end{equation*}
where ${{f^{eq}} = f^{eq}}({\xi};{\rho},{u})$, and ${\rho}$, ${u}$ is given by the analytical solution in Eq. \eqref{TV_AS}.

\begin{figure}
	\centering
	\subfigure[]{
		\label{fig:TV_T1}
		\includegraphics[width=2.7in,height=2.5in]{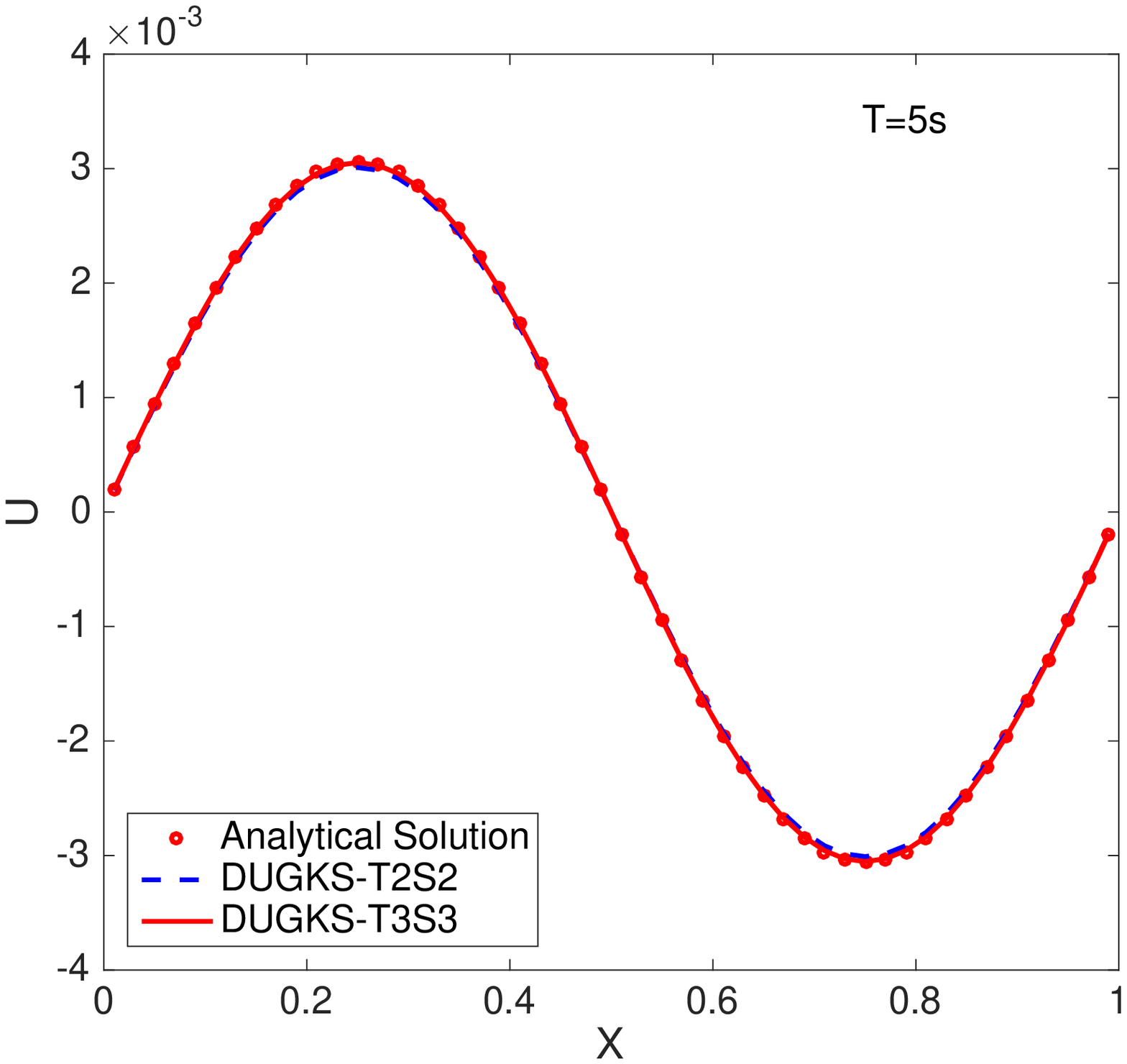}}
	\subfigure[]{
		\label{fig:TV_T5}
		\includegraphics[width=2.7in,height=2.5in]{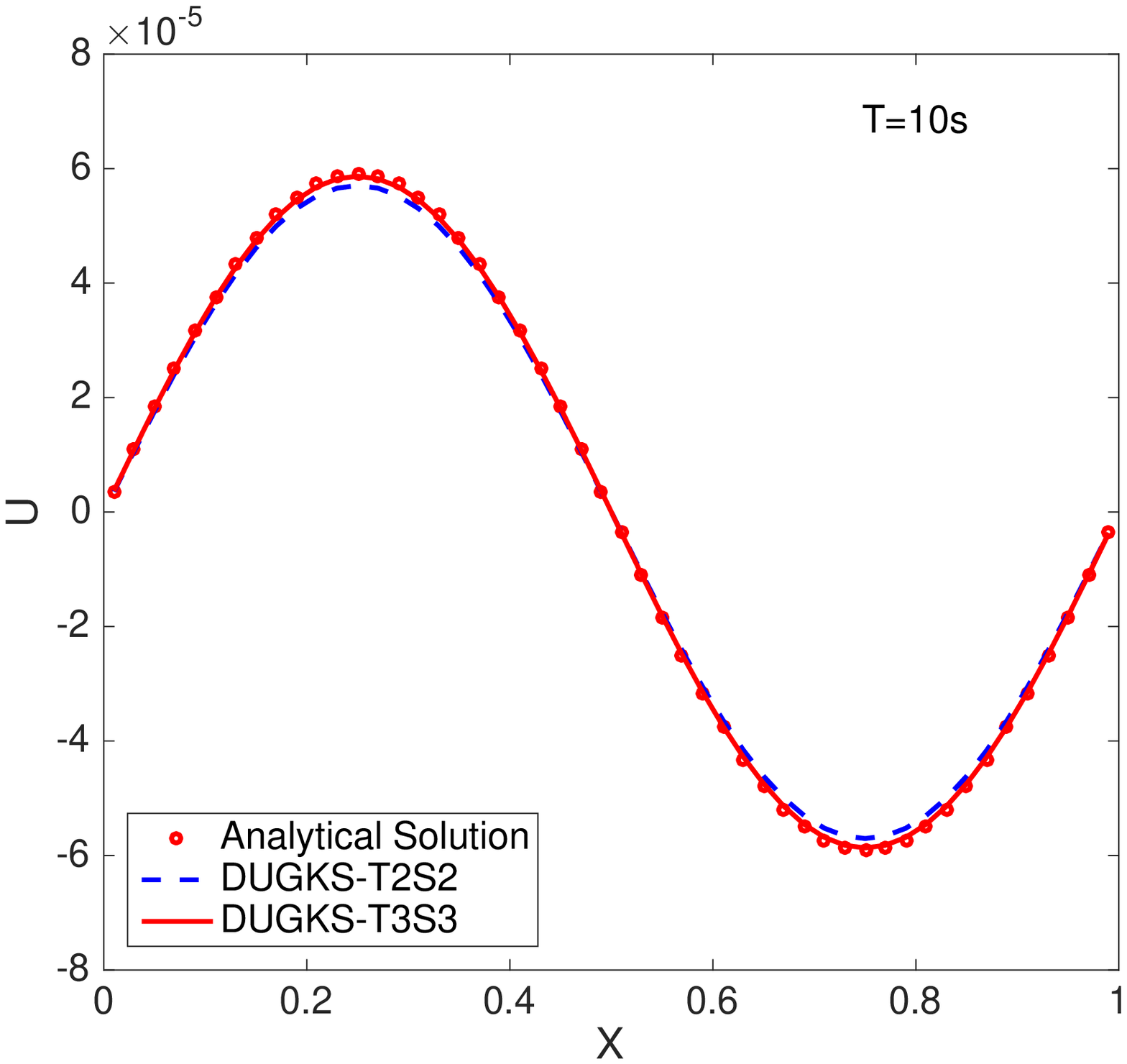}}
	\subfigure[]{
		\label{fig:TV_T10}
		\includegraphics[width=2.6in,height=2.5in]{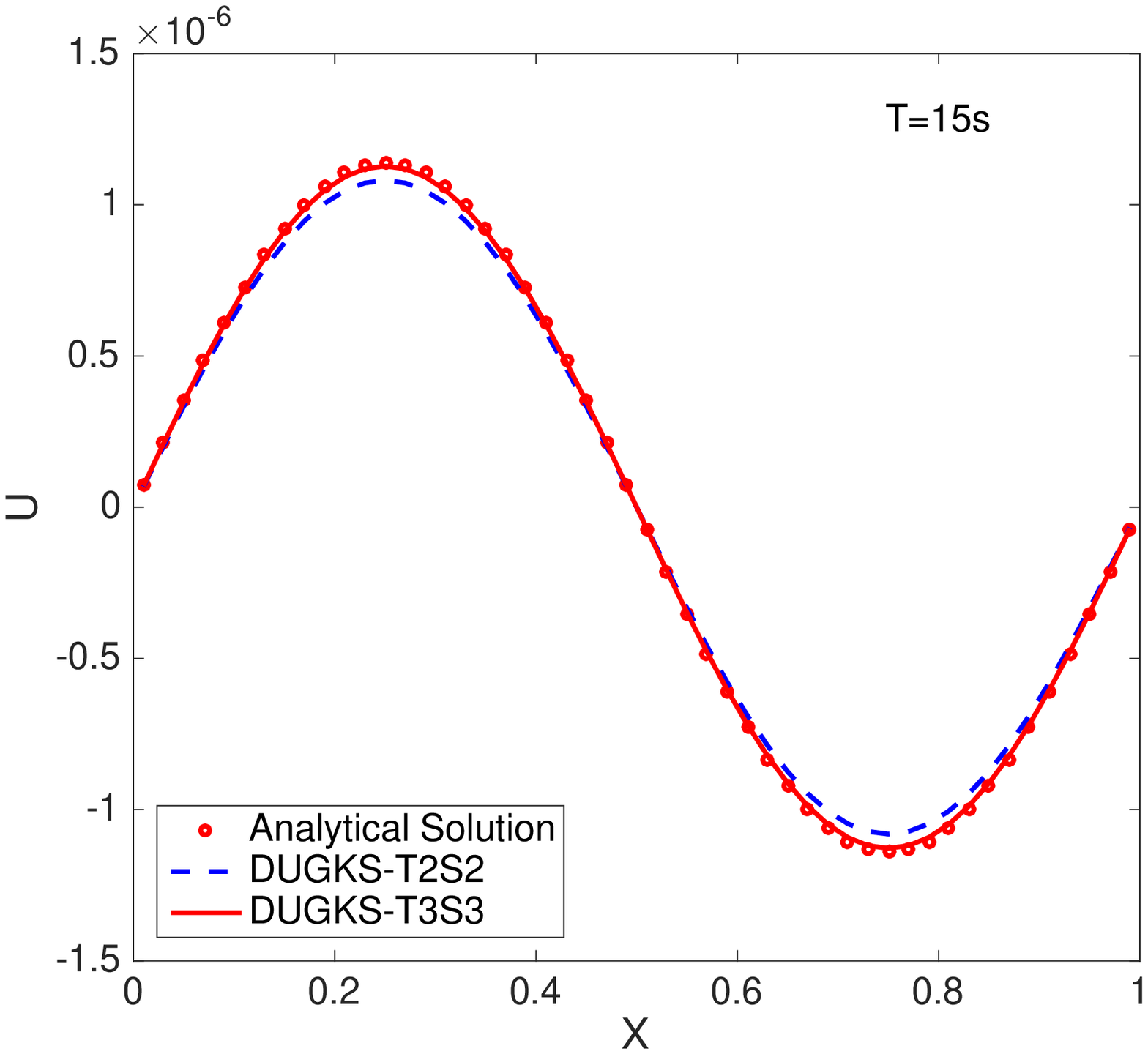}}
	\subfigure[]{
		\label{fig:TV_T15}
		\includegraphics[width=2.7in,height=2.5in]{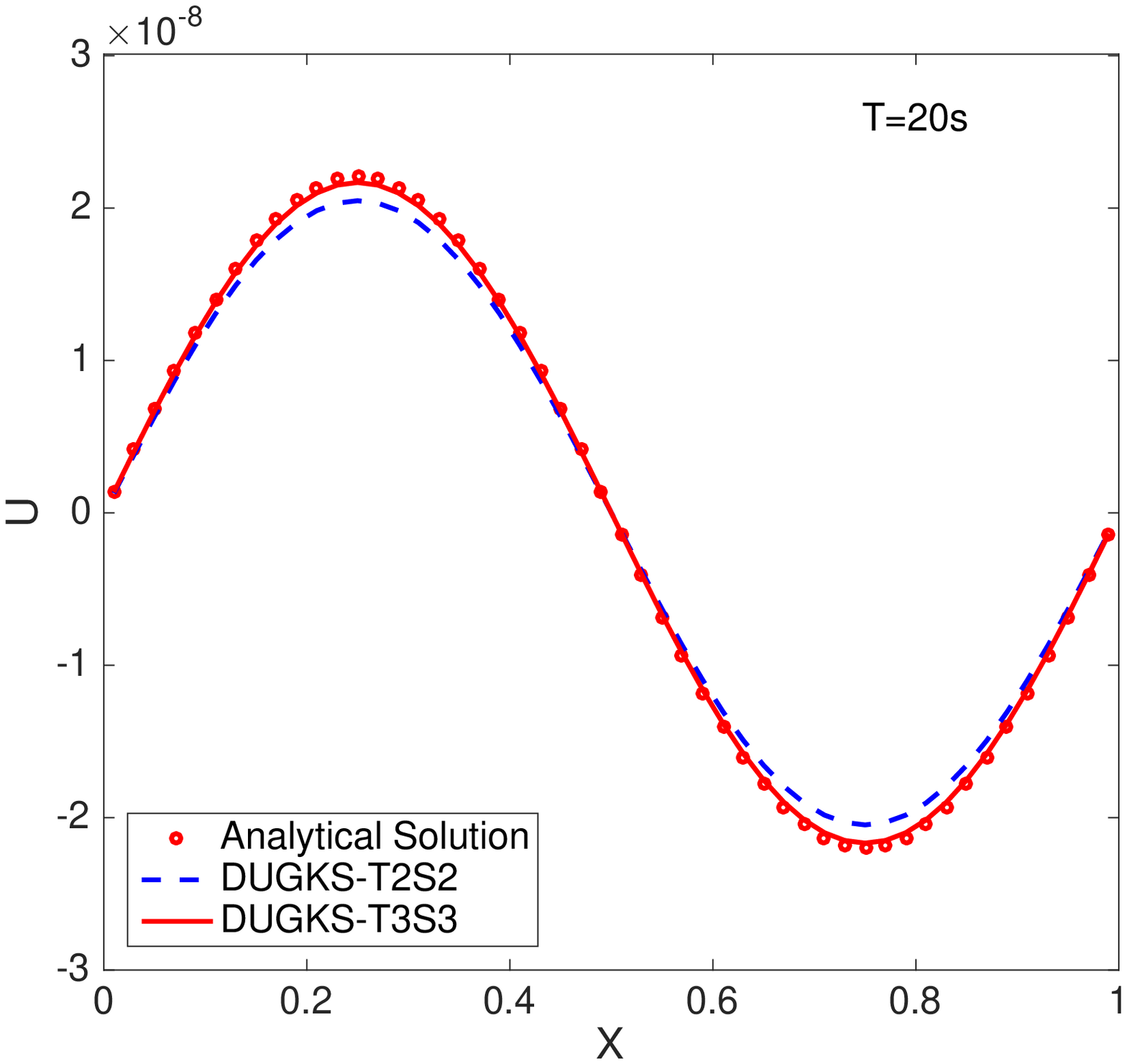}}
	\caption{Velocity profiles on $50 \times 50$ uniform mesh at different times.} 
	\label{fig:TV_T}
\end{figure}

The comparison of velocity profiles among DUGKS-T2S2, DUGKS-T3S3 and analytical solution at different moments are shown in Fig. \ref{fig:TV_T}. It can be observed that, at early moment (e.g. $t=5s$) both DUGKS-T2S2 and DUGKS-T3S3 are in good agreement with the analytical solution. However, more significant errors are observed in DUGKS-T2S2 as time marches. For instance, at $t = 20s$, the maximum deviation from the analytical solution in the results given by DUGKS-T2S2 is $2.8 \%$, while the maximum deviation in DUGKS-T3S3 is only $0.8 \%$.

In order to validate the convergence order of DUGKS-T3S3, we perform a set of simulations on different mesh sizes ($ N \times N, N = 32, 64, 96, 128, 160, 192$) and measured the $L_2$ global relative errors of velocity fields at $t=1s$,
\begin{equation}
E(\bm{u}) = \frac{{\sqrt {\sum\nolimits_{i,j} {{{\left| {\bm{u}_{ij} - \bm{u} '_{ij}} \right|}^2}} } }}{{\sqrt {\sum\nolimits_{i,j} {{{\left| {\bm{u} '_{ij}} \right|}^2}} } }}, 
\end{equation}
where  $\bm{u}_{ij} '$ is the analytical solution given by Eq. \eqref{TV_AS}.

In computation, the time step is maintained at $\Delta t = 1 \times 10^{-5}$. Parameters are chosen as $\frac{\Delta t}{\tau} = 4$, $\nu = 0.01$.

Table. \ref{table:EC_VP4}, presents the global relative errors and the convergence orders of various methods at different mesh sizes. It can be observed in the table that DUGKS-T2S2 model is approximately in the second order of accuracy. The DUGKS-T2S3 model, although having employed the third-order scheme in space discretization, is constrained by its second-order in temporal accuracy, due to which its overall accuracy is between the second- and the third-order. The DUGKS-T3S3 model proposed in this paper, is proven to be a “truly” third-order scheme. Both the time discretization and the space discretization of this model ensure the third-order of accuracy. Another notable issue is that, compared with DUGKS-T2S2 and DUGKS-T2S3, the relative error of DUGKS-T3S3 at the same mesh size is much smaller, especially on the refined meshes.

%\begin{center}
%	\begin{table}
%		\caption{Errors and convergence orders of steady velocity and pressure with $\frac{\Delta t}{\tau} = 4, RT=2000$.}
%		\begin{tabularx}{\textwidth}{l l X<{\centering} X<{\centering} X<{\centering} X<{\centering}}
%			\hline
%			\hline
%			Model & $N$         & $25$    & $50$  & $100$ & $200$ \\ 
%			\hline
%			
%			T2S2 & $E(\bm{u})$  & $1.426 \times 10^{-2}$ & $3.623 \times 10^{-3}$ & $9.130 \times 10^{-4}$ & $2.322 \times 10^{-4}$ \\
%			& order       & $-$  & $1.977$ & $1.989$  & $1.975$   \\
%			
%			T2S3 & $E(\bm{u})$  & $1.950 \times 10^{-2}$ & $2.124 \times 10^{-3}$ & $3.741 \times 10^{-4}$ & $5.475 \times 10^{-5}$ \\
%			& order       & $-$  & $3.198$ & $2.506$  & $2.773$   \\
%			
%			T3S3 & $E(\bm{u})$  & $1.568 \times 10^{-2}$ & $7.893 \times 10^{-4}$ & $8.018 \times 10^{-5}$ & $6.702 \times 10^{-6}$ \\
%			& order       & $-$  & $4.312$ & $3.300$  & $3.581$   \\
%			\hline
%			\hline
%		\end{tabularx}
%		\label{table:EC_VP4}
%	\end{table}
%\end{center}

\begin{center}
	\begin{table}
		\caption{Errors and convergence orders of steady velocity and pressure with $\frac{\Delta t}{\tau} = 4, dt=1 \times 10^{-5}$.}
		\begin{tabularx}{\textwidth}{l l X<{\centering} X<{\centering} X<{\centering} X<{\centering} X<{\centering} X<{\centering} }
			\hline
			\hline
			Model & $N$         & $32$    & $64$  & $96$ & $128$ & $160$ & $192$\\ 
			\hline
			
			T2S2 & $E(\bm{u})$  & $8.772 \times 10^{-3}$ & $2.215 \times 10^{-3}$ & $9.876 \times 10^{-4}$ & $5.570 \times 10^{-4}$  & $3.575 \times 10^{-4}$ & $2.490 \times 10^{-4}$\\
			& order       & $-$  & $1.985$ & $1.992$  & $1.991$ & $1.987$  & $1.982$   \\
			
			T2S3 & $E(\bm{u})$  & $9.669 \times 10^{-3}$ & $1.244 \times 10^{-3}$ & $4.669 \times 10^{-4}$ & $2.351 \times 10^{-4}$ & $1.357 \times 10^{-4}$ & $8.470 \times 10^{-5}$\\
			& order       & $-$  & $2.985$ & $2.417$  & $2.385$ & $2.462$  & $2.585$   \\
			
			T3S3 & $E(\bm{u})$  & $6.840 \times 10^{-3}$ & $3.989 \times 10^{-4}$ & $1.102 \times 10^{-4}$ & $4.866 \times 10^{-5}$ & $2.577 \times 10^{-5}$ & $1.480 \times 10^{-5}$ \\
			& order       & $-$  & $4.099$ & $3.173$  & $2.838$  & $2.847$  & $3.048$  \\
			\hline
			\hline
		\end{tabularx}
		\label{table:EC_VP4}
	\end{table}
\end{center}

%Secondly, the sizes of mesh are set to be $32, 64, 128$ and $256$. In Table. \ref{table:EC_VP3}, we may find the same conclusion as above. But the accuracy order of $T2S3$ increases to $3.201$ on the mesh of $256$. It is hard to explain.
%\begin{center}
%	\begin{table}
%		\caption{Errors and convergence orders of steady velocity and pressure with $\frac{\Delta t}{\tau} = 4, RT=2000$.}
%		\begin{tabularx}{\textwidth}{l l X<{\centering} X<{\centering} X<{\centering} X<{\centering}}
%			\hline
%			\hline
%			Model & $N$           & $32$    & $64$  & $128$ & $256$ \\ 
%			\hline
%			
%			T2S3 & $E(\bm{u})$  & $8.024 \times 10^{-3}$ & $1.126 \times 10^{-3}$ & $1.993 \times 10^{-4}$ & $2.167 \times 10^{-5}$ \\
%			& order       & $-$  & $2.834$ & $2.498$  & $3.201$   \\
%			
%			T3S3 & $E(\bm{u})$  & $5.057 \times 10^{-3}$ & $3.260 \times 10^{-4}$ & $3.748 \times 10^{-5}$ & $3.090 \times 10^{-6}$ \\
%			& order       & $-$  & $3.955$ & $3.121$  & $3.600$   \\
%			\hline
%			\hline
%		\end{tabularx}
%		\label{table:EC_VP3}
%	\end{table}
%\end{center}

\subsection{Doubly periodic shear layers}
The doubly periodic shear layers flow is employed to test the present method. The problem is initialized with the perturbation of two shear layers in a square domain and continues by the evolution of the shear layers due to the Kelvin-Helmholtz instability effect. Both numerical accuracy and stability are required to capture such delicate flow structures \cite{ hejranfar2017high, dellar2001bulk}. 

The computational domain in our simulation is a square $0 \le x,y \leq 1$. Periodic boundary conditions are implemented on all boundaries. And the initial velocity field is given as follows,
\begin{equation}
	\begin{split}
		{u_x} =& \left\{ {\begin{array}{*{20}{c}}
					{\tanh (k(y - 1/4)),y \le 1/2}\\
					{\tanh (k(3/4 - y)),y > 1/2}
			\end{array}} \right.\\
		{u_y} =& \delta \sin (2\pi (x + 1/4)),
	\end{split}
\end{equation}
where the parameters $k$ and $\delta$ are used to control the width of the shear layers and the magnitude of the initial perturbation, respectively. 
\begin{figure}
	\centering
%	\subfigure[]{
%		\label{fig:Shear_Flow_2nd_128_08}
%		\includegraphics[width=2.3in,height=2.4in]{2nd_128_t0_8}} 
%	\subfigure[]{
%		\label{fig:Shear_Flow_2nd_128_1}
%		\includegraphics[width=2.3in,height=2.4in]{2nd_128_t1}}
%	\subfigure[]{
%		\label{fig:Shear_Flow_3rd_128_08}
%		\includegraphics[width=2.3in,height=2.4in]{3rd_128_t0_8}}
%	\subfigure[]{
%		\label{fig:Shear_Flow_3rd_128_1}
%		\includegraphics[width=2.3in,height=2.4in]{3rd_128_t1}}
%	\subfigure[]{
%		\label{fig:Shear_Flow_2nd_256_08}
%		\includegraphics[width=2.3in,height=2.4in]{2nd_256_t0_8}}
%	\subfigure[]{
%		\label{fig:Shear_Flow_2nd_256_1}
%		\includegraphics[width=2.3in,height=2.4in]{2nd_256_t1}}
\subfigure[]{
	\label{fig:Shear_Flow_2nd_128}
	\includegraphics[width=4.9in,height=2.4in]{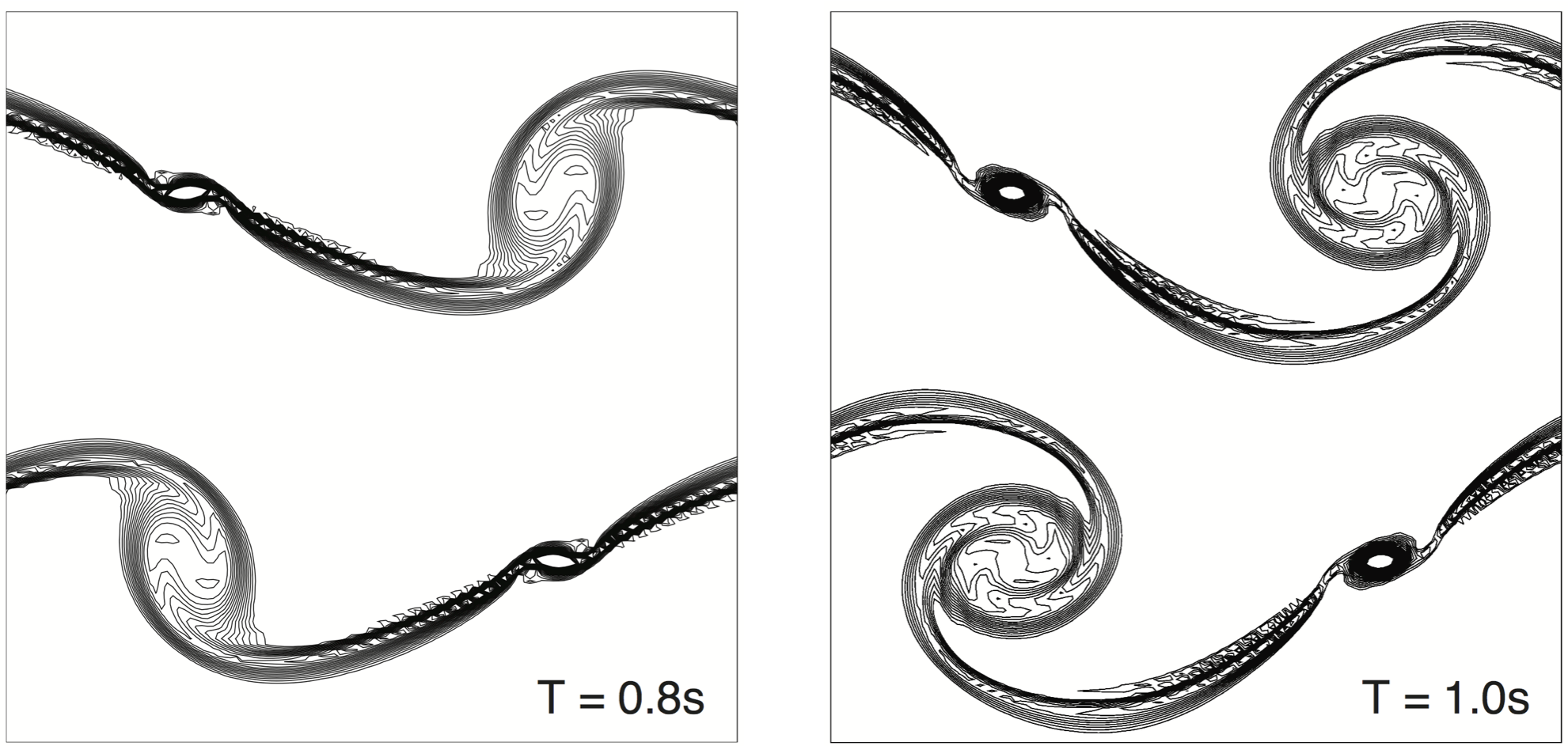}}
\subfigure[]{
	\label{fig:Shear_Flow_3rd_128}
	\includegraphics[width=4.9in,height=2.4in]{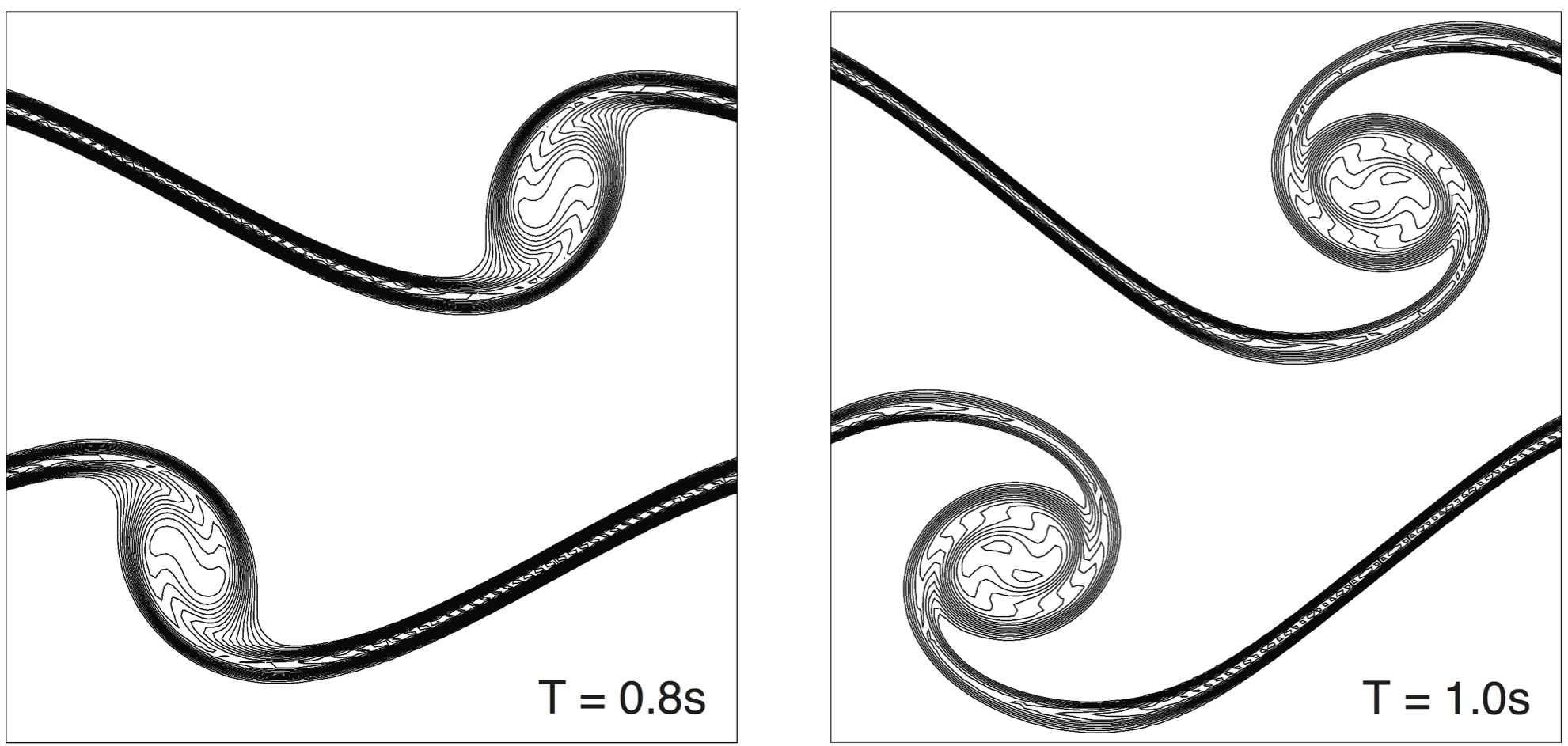}}
\subfigure[]{
	\label{fig:Shear_Flow_2nd_256}
	\includegraphics[width=4.9in,height=2.4in]{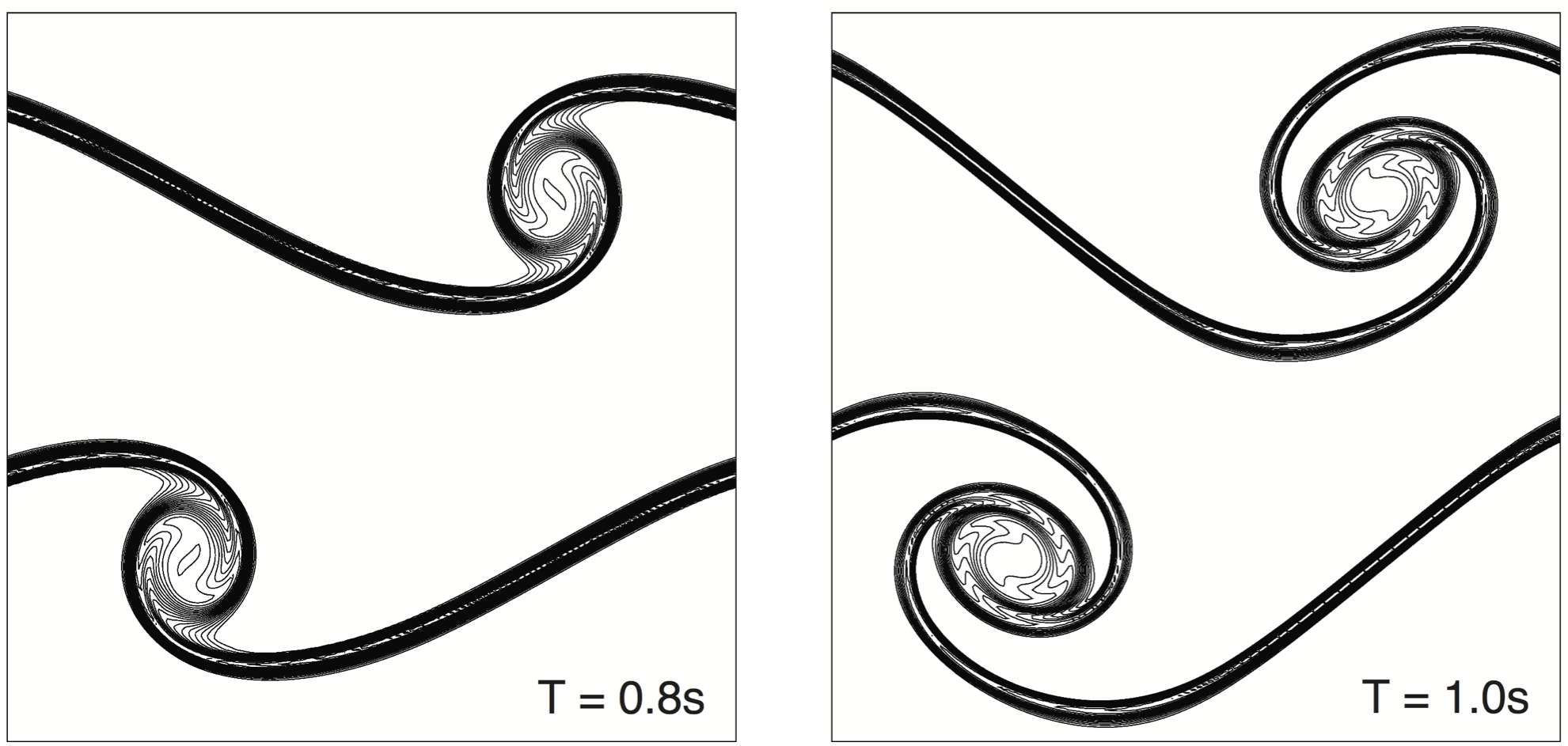}} 
	\caption{Contours of vorticity at $T=0.8s$ and $T=1.0s$. (a) DUGKS-T2S2 on the mesh size of $128 \times 128$; (b) DUGKS-T3S3 on the mesh size of $128 \times 128$; (c) DUGKS-T2S2 on the mesh size of $256 \times 256$.} 
	\label{fig:Shear_Flow_Re10000}
\end{figure}
Simulations are carried out with $k = 80$, $\delta = 0.05$ and $\nu = 1 \times 10^{-4}$, using DUGKS-T3S3 method on the mesh size of $128 \times 128$ and using DUGKS-T2S2 method on the mesh sizes of $128 \times 128$ and $256 \times 256$. 

Fig. \ref{fig:Shear_Flow_Re10000} presents the vorticity contours obtained by DUGKS-T2S2 and DUGKS-T3S3 at different time and mesh sizes. It turns out that DUGKS-T3S3 can give smooth and stable results at coarse mesh ($128 \times 128$), which indicates its nice numerical accuracy and stability. The DUGKS-T2S2 model, on the other hand, shows poorer performance, which is reflected as the emerging spurious vortices. To successfully capture the delicate vortex structure, DUGKS-T2S2 model requires finer mesh ($256 \times 256$).

\subsection{Continuum lid-driven cavity flow} 
\label{sec:LDF}
Two-dimensional continuum lid-driven cavity flow (LDF) is a classic benchmark test. In this part, we use this example to validate the present DUGKS-T3S3 model in low-speed continuum flow. The physical configuration of the problem is a two-dimensional unit square with a lid moving along the horizontal direction in a constant velocity $U$. In our simulations, the Mach number is set to be  $Ma=0.15$. and the Reynolds number is defined as $Re = UL/\nu$ where $L = 1.0$ represents the length of the cavity, and $U = 1.0$. The BB scheme is implemented on all boundaries. Besides, the following criterion is used to ensure that the problem reaches steady state,

\[\frac{{\sqrt {\sum\nolimits_{i,j} {{{\left| {u_{ij}^{(n + 1000)} - u_{ij}^{(n)}} \right|}^2}} } }}{{\sqrt {\sum\nolimits_{i,j} {{{\left| {u_{ij}^{(n + 1000)}} \right|}^2}} } }} \le {10^{ - 6}},\]
where $u_{ij}^n = u\left( {{x_i},{y_j},n\Delta t} \right)$. 

\begin{figure}
	\centering
	\subfigure[]{
		\label{fig:LDF_Period_Re12000_1}
		\includegraphics[width=2in,height=1.9in]{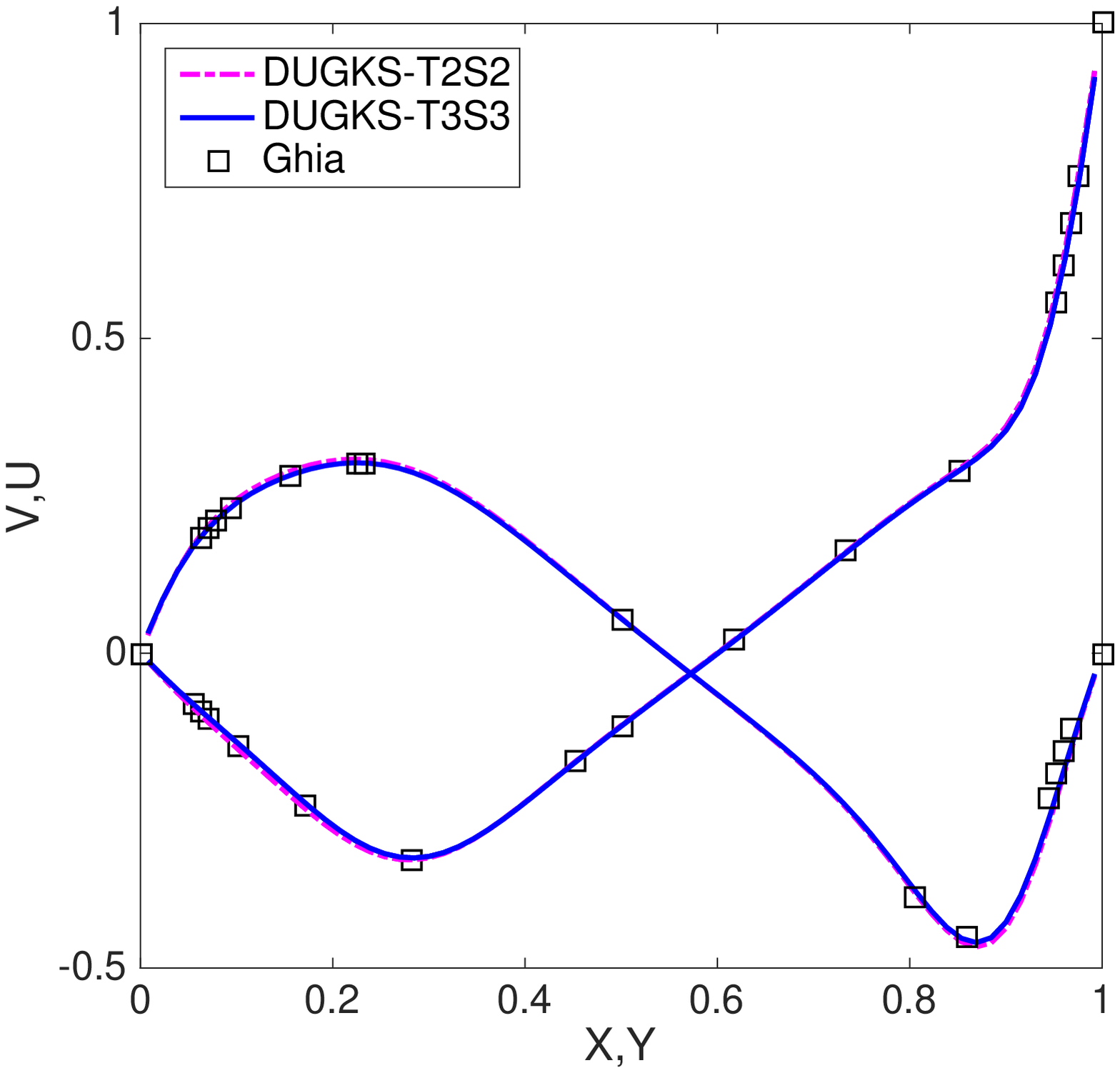}}
	\subfigure[]{
		\label{fig:LDF_Period_Re12000_2}
		\includegraphics[width=2in,height=1.9in]{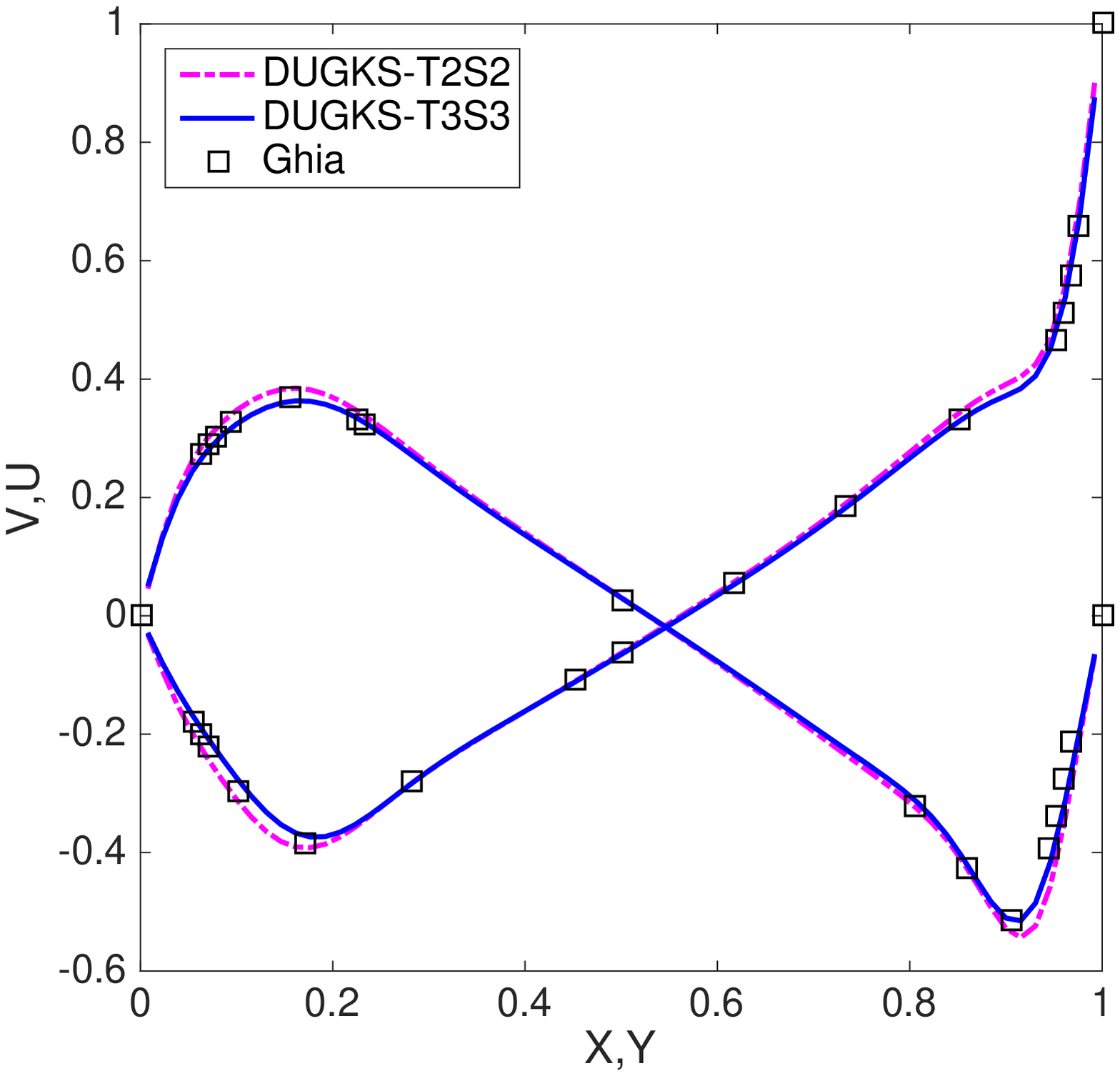}}
	\subfigure[]{
		\label{fig:LDF_Period_Re12000_3}
		\includegraphics[width=2in,height=1.9in]{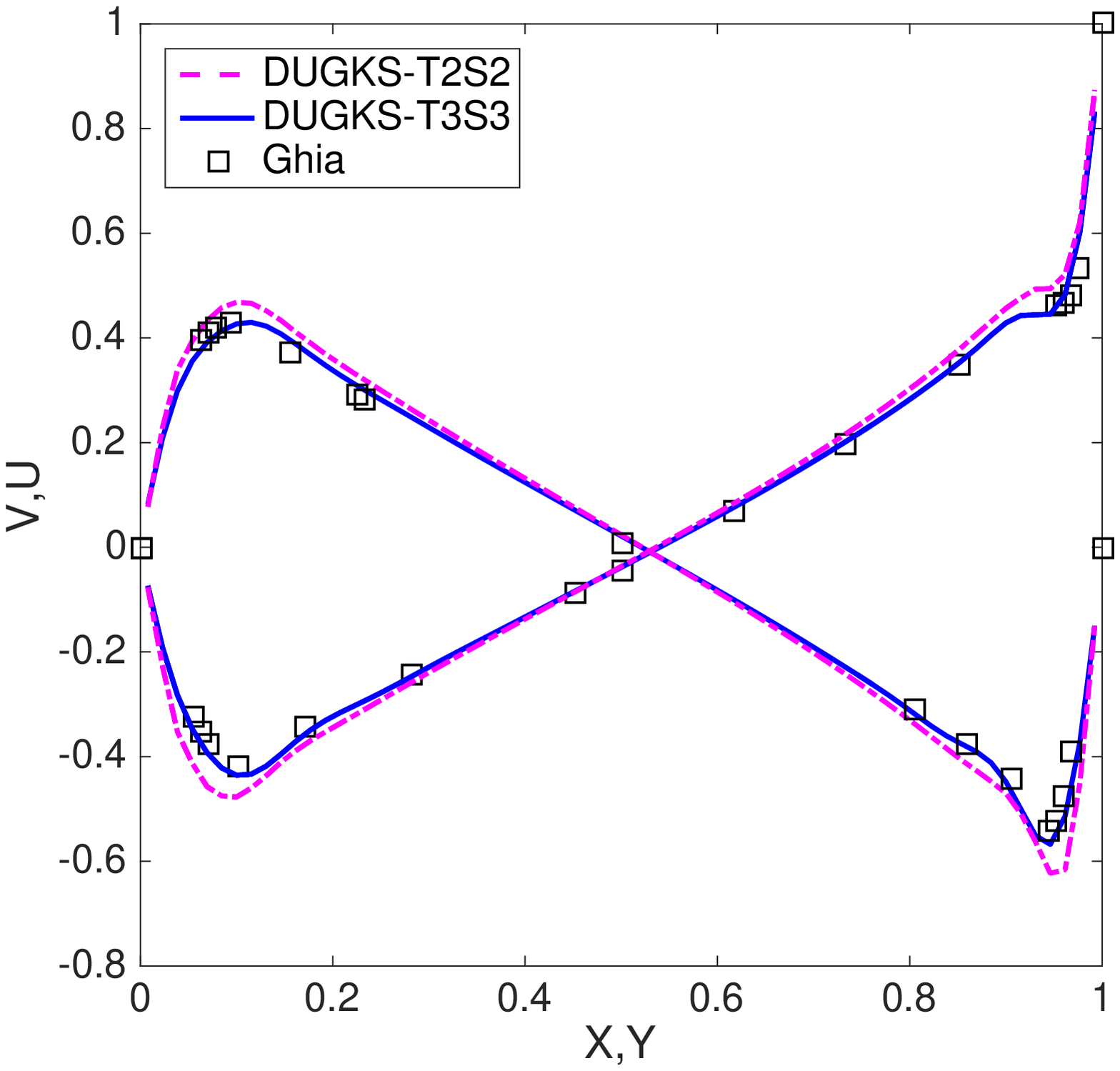}}
	\caption{Velocity profiles at $Re=400, 1000$ and $3200$ on $65 \times 65$ uniform mesh. ($\circ$) is reference data. (a) $U$ along the vertical centerline; (b) $V$ along the horizontal centerline.} 
	\label{fig:LDF_Period_Re12000}
\end{figure}

\begin{figure}
	\centering
%	\subfigure[]{
%		\label{fig:LDF_T2S2_S}
%		\includegraphics[width=2.7in,height=2.5in]{T2S2_LDF_Re3200_S}}
	\subfigure[]{
		\label{fig:LDF_T2S2_P}
		\includegraphics[width=2.7in,height=2.5in]{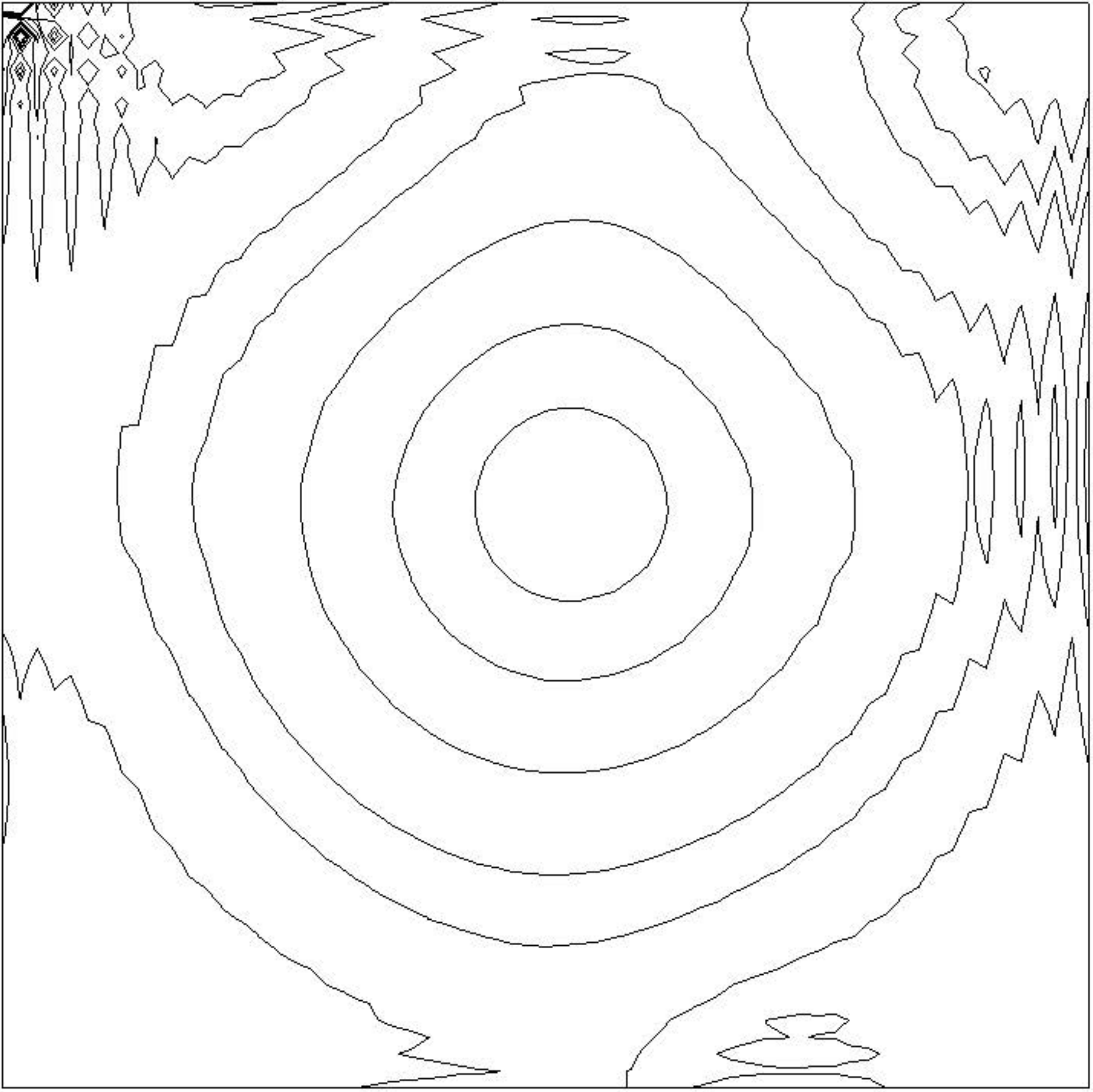}}
%	\subfigure[]{
%		\label{fig:LDF_T3S3_S}
%		\includegraphics[width=2.6in,height=2.5in]{T3S3_LDF_Re3200_S}}
	\subfigure[]{
		\label{fig:LDF_T3S3_P}
		\includegraphics[width=2.7in,height=2.5in]{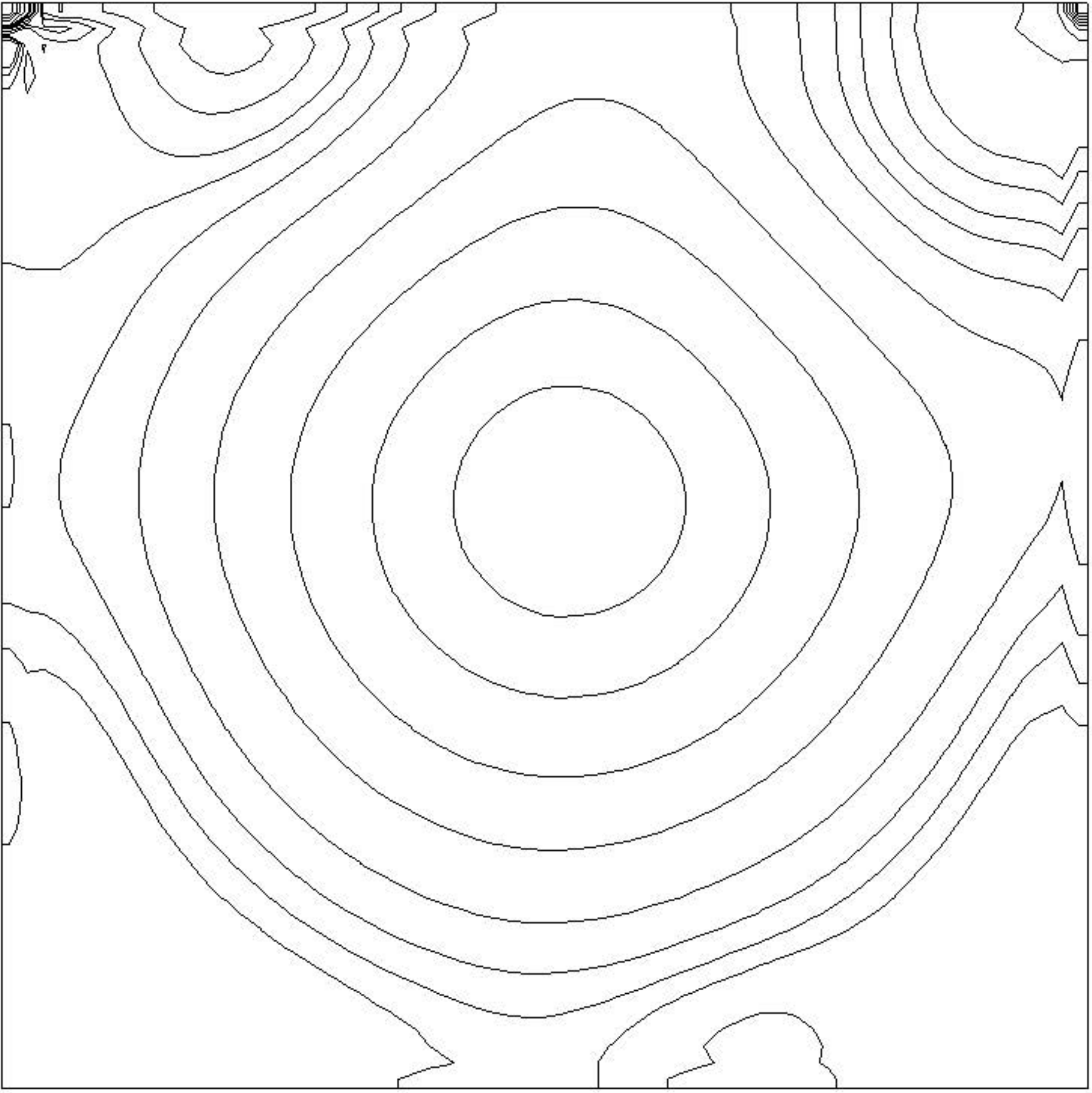}}
	\caption{Pressure contours at $Re=3200$ on $65 \times 65$ uniform mesh. (a) is obtained by T2S2 method. (b) is obtained by T3S3 method.} 
	\label{fig:LDF_Re3200_S_P}
\end{figure}

Simulations are carried out on the uniform mesh size of $65 \times 65$ and at various Reynolds number of $Re = 400, 1000$ and $3200$. Quantitative comparisons are offered in Fig. \ref{fig:LDF_Period_Re12000}, which compares the velocity distribution along the centerlines with reference data. As shown in these figure, when the Reynolds numbers are low, e.g. $Re = 400$ and $1000$, the numerical results of T2S2 and T3S3 are both in good agreement with the reference data. However, as the Reynolds number increased, the results of T3S3 agree better with the reference date, especially in the boundary layer region. The difference originates from the higher order of accuracy in DUGKS-T3S3 model. Generally, DUGKS-T3S3 model requires smaller mesh size than DUGKS-T2S2 model to get converged results.
Pressure contours obtained by DUGKS-T2S2 and DUGKS-T3S3 at $Re = 3200$ are shown in Fig. \ref{fig:LDF_Re3200_S_P}. The result of DUGKS-T2S2 contains strong oscillations near the boundaries, while the solution of DUGKS-T3S3 is more smooth, which indicates that DUGKS-T3S3 method is more accurate.

\begin{figure}
	\centering
	\includegraphics[width=2.7in,height=2.5in]{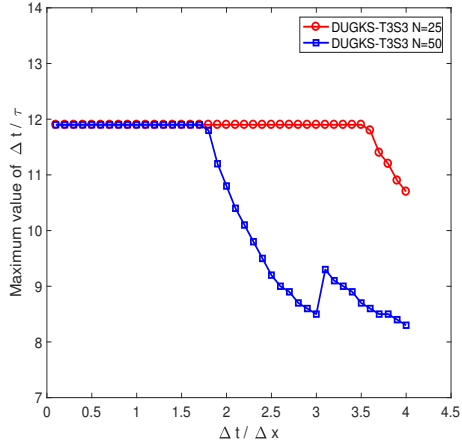}
	\caption{The maximum values of $\Delta t / \tau$ for stable computations of the LDF on $25 \times 25$ and $50 \times 50$ meshes.} 
	\label{fig:LDF_Re100}
\end{figure}

In addition, the limitation of $\Delta t/\tau$ of the present DUGKS-T3S3 model is also studied to validate its numerical stability. Most of the Boltzmann equation based high-order methods are constrained by the limitation of $\Delta t/\tau < 2$. Even by employing more stable scheme in time and space discretization, the limitation can just be slightly lifted to $\Delta t/\tau < 5$ in most cases \cite{hejranfar2017high}, and the ratio of $\Delta t/\Delta x$ should be kept in a small value. As a comparison, Fig. \ref{fig:LDF_Re100} shows the maximum values of $\Delta t/\tau$ of DUGKS-T3S3 method in the example of LDF at $Re = 100$ on $25 \times 25$ and $50 \times 50$ meshes. It can be observed that the limitation of DUGKS-T3S3 is around $\Delta t/\tau < 12$, which is consistent with the analysis in Sec. \ref{sec:time_accuracy}. Besides, the stability boundary does not vary with the ratio of $\Delta t$ and $\Delta x$ until $\Delta t/\Delta x > 2$, which indicates that larger time step can be adopted in DUGKS-T3S3 method and less computation time will be consumed. Noted that, although the stability boundary of DUGKS-T3S3 is smaller than DUGKS-T2S2 \cite{zhu2017performance}, it still has better numerical stability than the Boltzmann equation based high-order methods \cite{hejranfar2017high,  su2015parallel}. From a numerical perspective, it is also normal that certain part of numerical stability has to be traded when uprising the numerical accuracy.

\subsection{Micro-cavity flow} 
\label{sec:micro_cavity}
In this case, the DUGKS-T3S3 is applied to simulate the micro-cavity flow, which has been studied by DSMC \cite{john2010investigation, john2011effects, mohammadzadeh2013parallel}, UGKS \cite{huang2012unified} and DVM \cite{yang2016numerical, yang2017comparative} methods. The accuracy and efficiency of present method for rarefied flow will be tested in this example. Different from continuum flow, the distribution function become highly irregular in the micro-cavity flow. The Gaussian quadrature method is no longer a proper choice as the distribution function will be span over the entire velocity space. Instead, the Newton-Cotes quadrature with sufficient discrete velocity points is employed to recover the irregular distribution function. Besides, the Maxwellian distribution can still be approximated by its Taylor expansions Eq. \eqref{maxwellian_expansion}, since the speed of micro-cavity flow is supposed to be low ($Ma << 1$).

The micro-cavity flows with different Knudsen numbers are simulated to validate the accuracy and efficiency of the present scheme. In our simulations, the velocity of top wall is given as $u_w / \sqrt{\gamma RT} = 0.16$. The particle velocity space $[-4 \sqrt{2RT},4 \sqrt{2RT}] \times [-4 \sqrt{2RT},4 \sqrt{2RT}]$ is discretized by $101 \times 101$ uniformly distributed points, on which Newton-Cotes quadrature is adopted. The computational domain is divided into $40 \times 40$ uniform cells uniformly. And the DS scheme \cite{guo2013discrete} is implemented as the boundary condition. The criterion of convergence defined in Sec. \ref{sec:LDF} is also employed in this case.

\begin{figure}
	\centering
	\subfigure[]{
		\label{fig:MLDF_T3S3_Kn0075}
		\includegraphics[width=2.7in,height=2.5in]{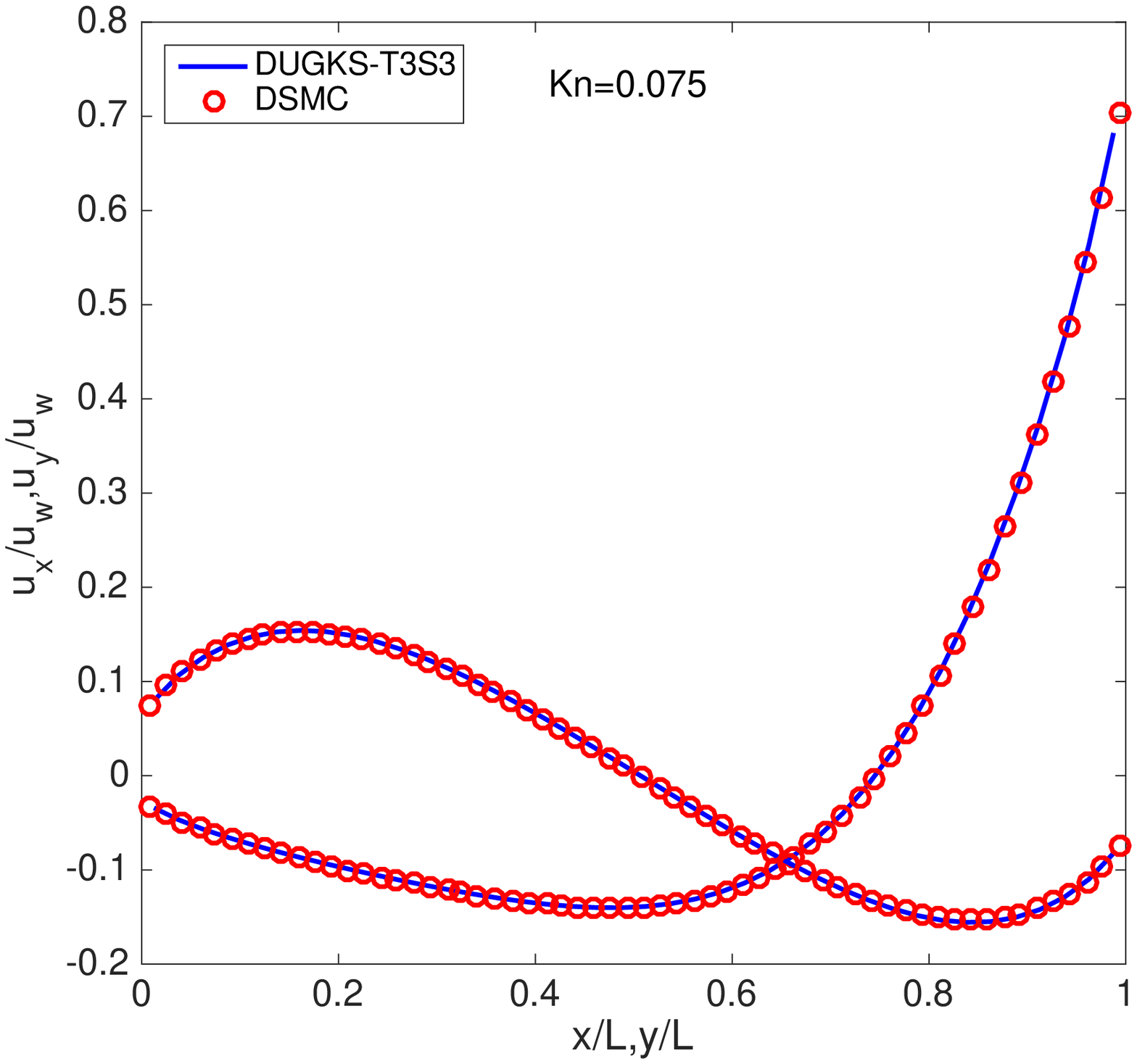}}
	\subfigure[]{
		\label{fig:MLDF_T3S3_Kn1}
		\includegraphics[width=2.7in,height=2.5in]{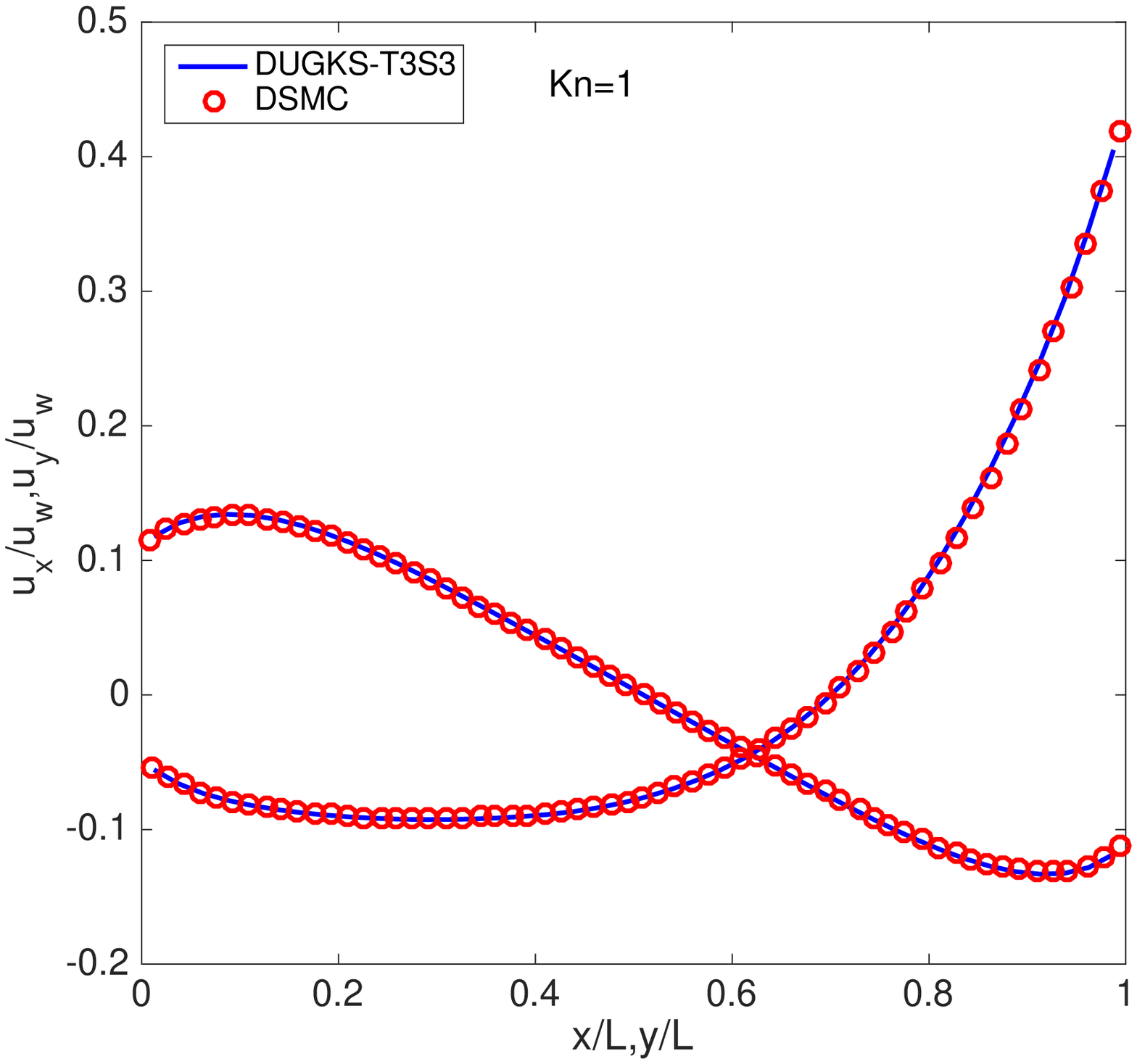}}
	\subfigure[]{
		\label{fig:MLDF_T3S3_Kn2}
		\includegraphics[width=2.7in,height=2.5in]{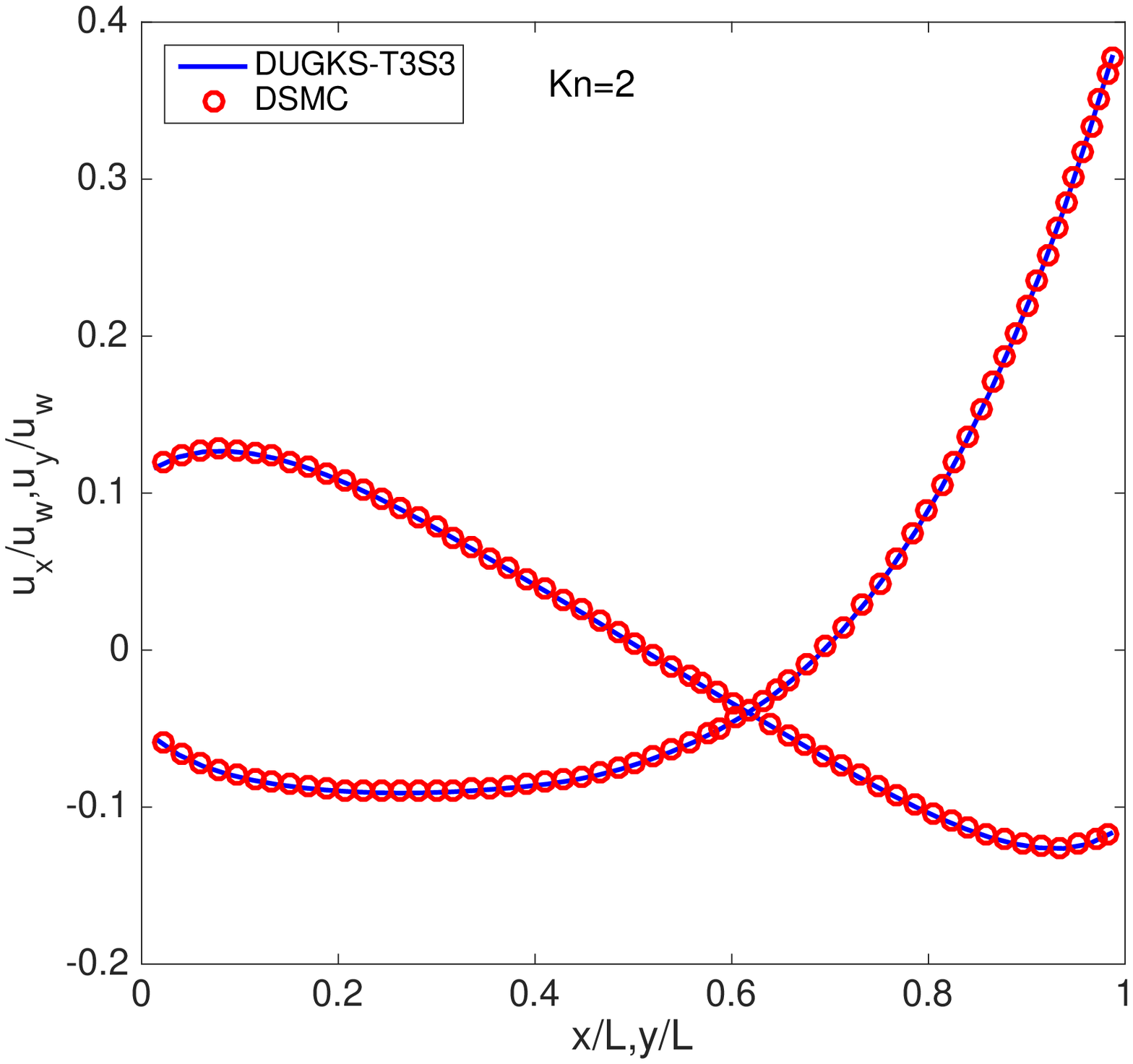}}
	\subfigure[]{
		\label{fig:MLDF_T3S3_Kn10}
		\includegraphics[width=2.7in,height=2.5in]{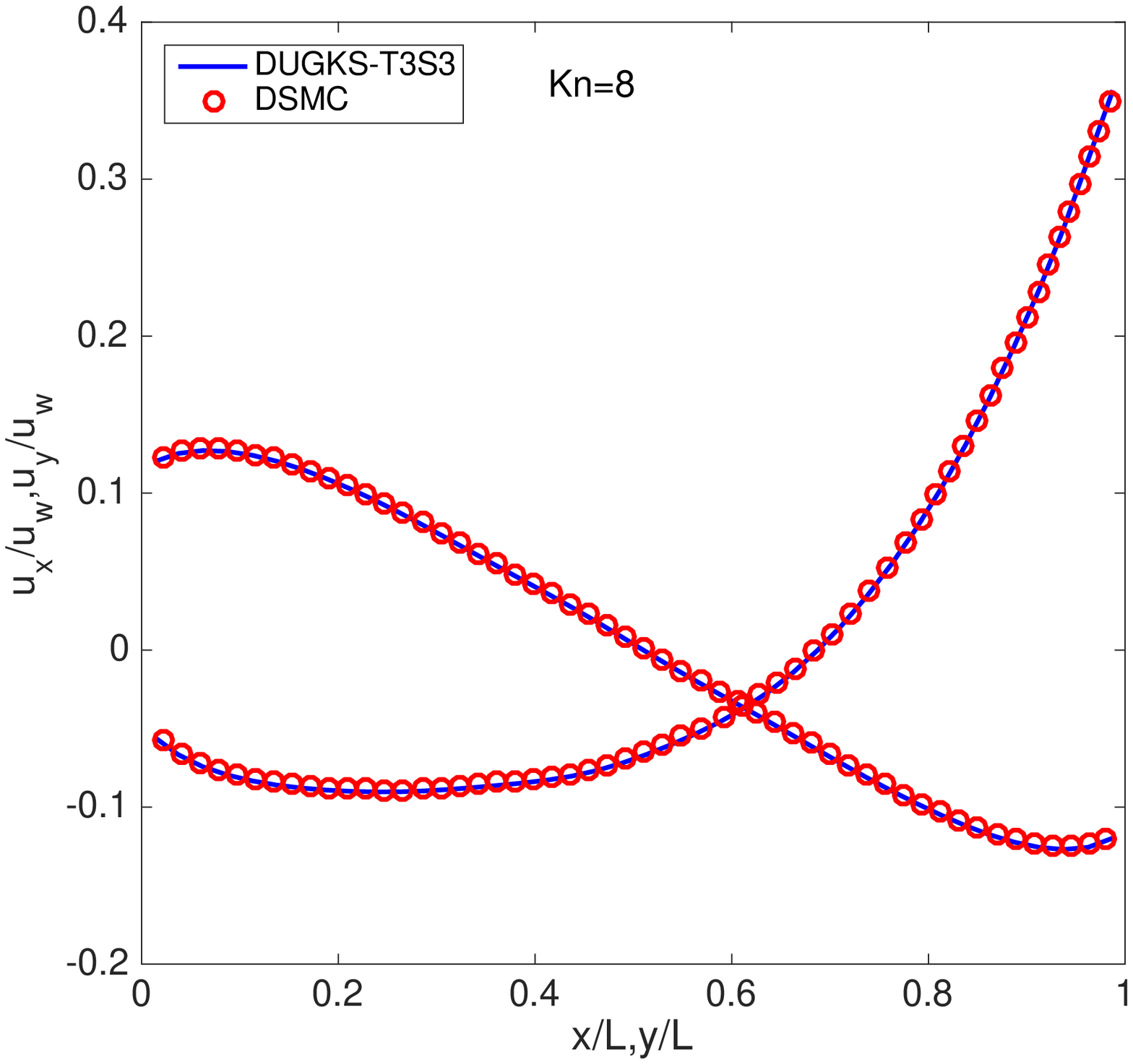}}
	\caption{Velocity profiles along the centerlines at $Kn=0.075, 1, 2$ and $8$ on $40 \times 40$ uniform mesh. ($\circ$) is reference data. (a) $U$ along the vertical center line; (b) $V$ along the horizontal center line.} 
	\label{fig:MLDF_Kn}
\end{figure}

Simulations are carried out at different Knudsen number of $Kn = 0.075, 1, 2$ and $8$, with $RT=0.5$ and $CFL = 0.25$. In Fig. \ref{fig:MLDF_Kn}, the velocity profiles along the cavity centerlines are shown and compared with the DSMC benchmark data \cite{john2010investigation, huang2012unified}. It can be observed that the results of DUGKS-T3S3 are in good agreement with the DSMC benchmarks, which indicates that the DUGKS-T3S3 is accurate in simulating rarefied flow, as well as the continuum flow.

\begin{figure}
	\centering
	\includegraphics[width=2.7in,height=2.5in]{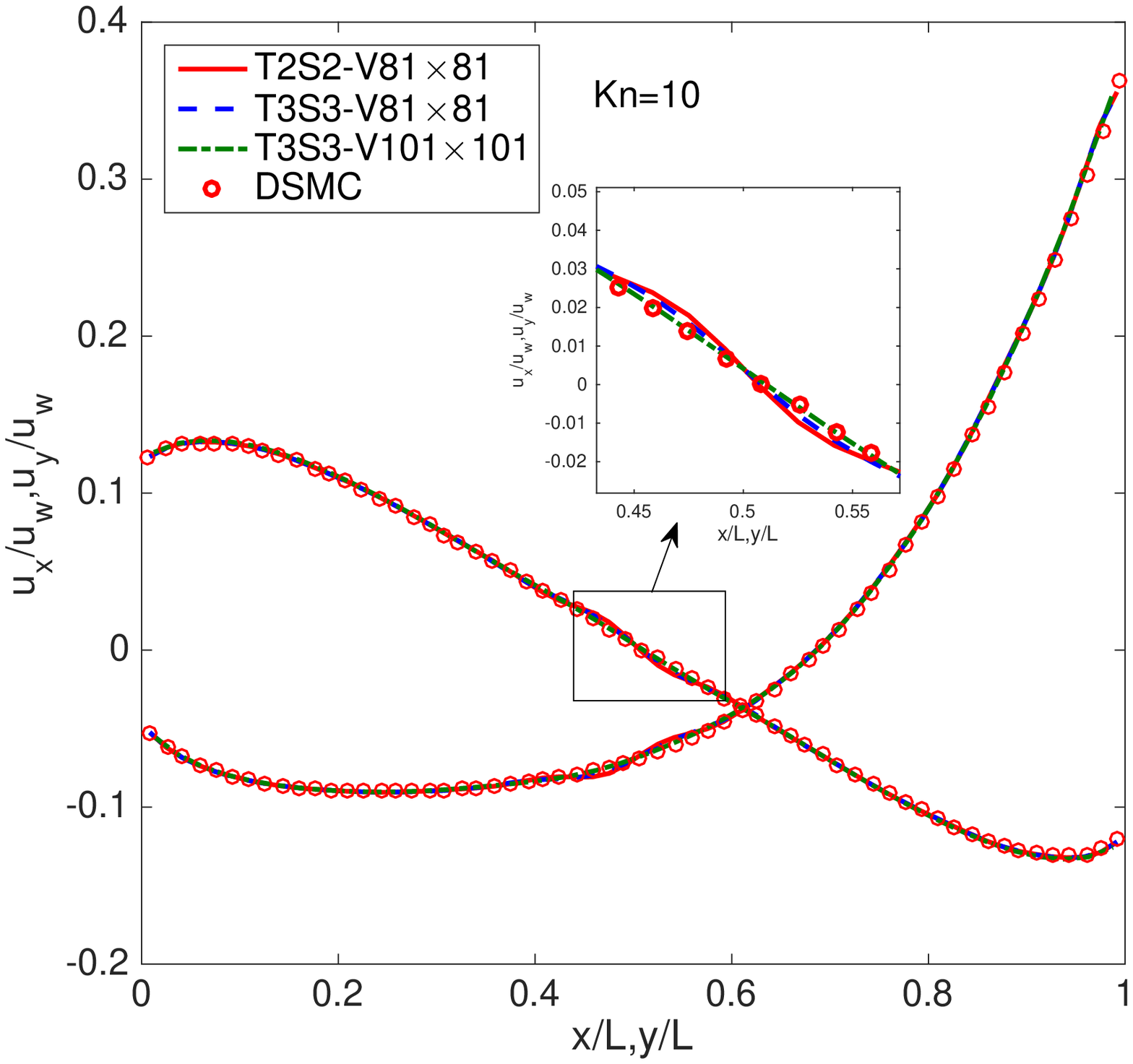}
	\caption{The velocity profiles of T2S2 and T3S3 methods at $Kn = 10$ on $60 \times 60$ uniform mesh with $CFL = 0.25$.} 
	\label{fig:MLDF_Period_Re2000}
\end{figure}

The second test is carried out at $Kn = 10$ and $CFL=0.25$, and $60 \times 60$ uniform mesh is adopted. Velocity profiles obtained by DUGKS-T2S2 and the present DUGKS-T3S3 are presented in Fig. \ref{fig:MLDF_Period_Re2000}. The zoom-in view reveals that, with $81 \times 81$ points particle velocities, both T2S2 model and T3S3 model show oscillations. Such oscillation is named as the “ray effect”, which is caused by the insufficient accuracy in integration over the velocity space. The magnitude of oscillation of T3S3 model is slightly smaller than that of T2S2 model, and can be further suppressed by refining the number of discrete particle velocity used in computation.

\begin{figure}
	\centering
	\includegraphics[width=2.7in,height=2.5in]{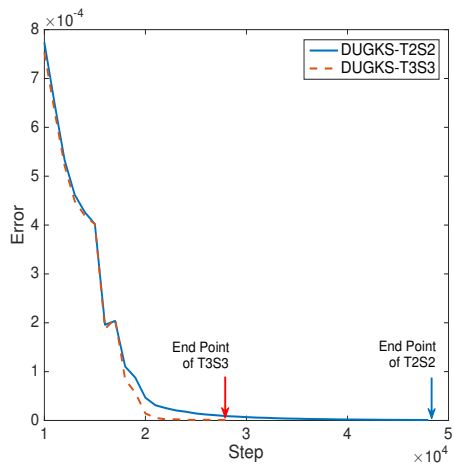}
	\caption{Error history of T2S2 and T3S3 methods at $Kn = 10$ on $60 \times 60$ uniform mesh with $CFL = 0.25$.} 
	\label{fig:MLDF_Period_Re200}
\end{figure}

Comparison of computational efforts between DUGKS-T2S2 and -T3S3 methods is shown in Table. \ref{my-label}. The computation time (per $100$ computation steps) of T3S3 is about 2 times of T2S2. However, the time history of relative error shown in Fig. \ref{fig:MLDF_Period_Re200} indicates that T3S3 method converges more quickly than T2S2, which means that T3S3 needs fewer steps to reach the convergence criteria. Actually, the time consumed by T3S3 model to trigger the convergence criteria is approximately the same ($1.05$ times) as the time used by T2S2 model. Table. \ref{my-label} also shows the cost of virtual memories of DUGKS-T2S2 and -T3S3 methods. The memory cost of T3S3 is about 1.3 times of T2S2. Analytically, the memory-consuming of T3S3 should be around $\frac{4N + 3}{3N + 3}$ times the cost of T2S2, where $N$ is the number of discrete velocities. One should be reminded that the above comparisons are made on the same mesh sizes. Since T3S3 model has higher order of accuracy than T2S2 model, smaller mesh size is required by T3S3 model to get converged results. Therefore, to obtain results in comparable accuracy, T3S3 model is definitely more competitive than T2S2 model in terms of computational efforts. 

\begin{figure}
	\centering
	\includegraphics[width=2.7in,height=2.5in]{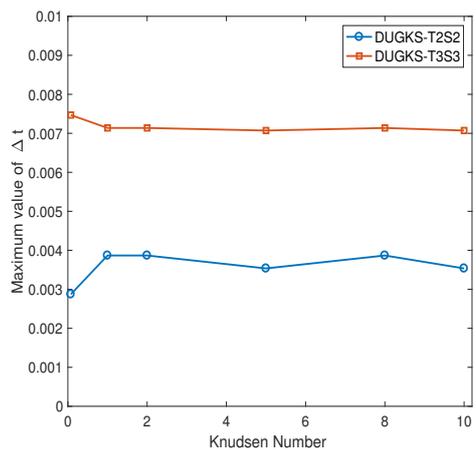}
	\caption{The maximum time-step size for stable computations of DUGKS-T2S2 and DUGKS-T3S3.} 
	\label{fig:Max_dt}
\end{figure}

\begin{table}[]
	\centering
	\caption{Comparison of time-consuming (per 100 computation steps) and memory-consuming.}
	\label{my-label}
	\begin{tabularx}{\textwidth}{l X<{\centering} X<{\centering} X<{\centering} X<{\centering}}
		\hline
		\hline
		\multicolumn{2}{l}{Mesh }                                             & $20 \times 20 $                      & $40  \times 40  $                    & $60   \times 60$                     \\
		\hline
		\multicolumn{1}{c}{\multirow{2}{*}{T3S3}} & \multicolumn{1}{c}{Mem (Mb)} & 19                       & 56                       & 116                      \\
		\multicolumn{1}{c}{}                      & \multicolumn{1}{c}{Time (s)} & 68.57                    & 277.89                   & 941.71                   \\
		\hline
		\multirow{2}{*}{T2S2}                     & Mem (Mb)                     & \multicolumn{1}{c}{16}   & \multicolumn{1}{c}{43}   & \multicolumn{1}{c}{93}   \\
		& \multicolumn{1}{c}{Time (s)} & 27.58                    & 167.28                   & 462.43                   \\
		\hline
		\multirow{2}{*}{Ratio}                    & Mem                          & \multicolumn{1}{c}{1.19} & \multicolumn{1}{c}{1.30} & \multicolumn{1}{c}{1.25} \\
		& Time                         & \multicolumn{1}{c}{2.49} & \multicolumn{1}{c}{1.66} & \multicolumn{1}{c}{2.04} \\
		\hline
		\hline
	\end{tabularx}
\end{table}
The maximum time-step size ($\Delta t_{max}$) for stable computations of DUGKS-T2S2 and DUGKS-T3S3 with same parameters is then investigated. Fig. \ref{fig:Max_dt} shows that $\Delta t_{max}$ of DUGKS-T3S3 is larger than DUGKS-T2S2 at all tested Knudsen numbers. The ratio of maximum time-step size of DUGKS-T3S3 and DUGKS-T2S2 is from $1.85$ to $2.60$. It implies that, compared with DUGKS-T2S2, DUGKS-T3S3 method shows better stability in time marching and can use larger time step, due to which less computation steps are needed by DUGKS-T3S3 to reach same physical time. 

\section{Conclusions}
\label{sec:conclusions}
In this paper, a third-order time-accurate discretization scheme for BE is proposed. Together with a third-order DUGKS reconstruction scheme, a DUGKS is developed for continuum and rarefied flows with third order accuracy in both time and space. From the analysis of BE using TFTD method, the original second-order DUGKS can also be proven as a special case of the present method. Different from the time splitting based kinetic methods, the present method is free from the constrain of the second-order of accuracy in the splitting process. Compared with the high-order kinetic methods that are based on LBE, the stability condition of $\Delta t / \tau$ is also improved to $12$ in the present method. The third order DUGKS has been validated by several classical problems including both continuum and rarefied cases. In all cases, the numerical solutions are in good agreement with the analytical results or the benchmark data. With these test cases, the third-order accuracy of the present method is validated, and the adoption of larger time-steps is also tested. Additionally, compared with the second-order DUGKS, the present third-order DUGKS consumes less computational time and virtual memories to get comparable results

\section*{Acknowledgments}
We gratefully acknowledge Dr. Liming Yang and Dr. Zhenhua Chai for their useful discussions during this work. This work is financially supported by the National Natural Science Foundation of China (Grants No. 51576079).

%%%% Bibliography  %%%%%%%%%%
\bibliographystyle{elsarticle-num}
%\biboptions{authoryear}
\bibliography{myrefs}

\end{document}